\documentclass[aps,prl,reprint,twocolumn,preprintnumbers,floatfix,nofootinbib]{revtex4-1}
\usepackage[utf8]{inputenc}
\usepackage[dvips]{graphicx}
\usepackage{xcolor}
\usepackage{color}
\usepackage{relsize}
\usepackage{graphics}
\usepackage{epstopdf}
\usepackage{hyperref}
\usepackage{mathrsfs}
\usepackage{gensymb}
\usepackage{amssymb}
\usepackage[caption=false]{subfig}

\usepackage[normalem]{ ulem }
\usepackage{soul}
\usepackage{amsthm}
\usepackage{amsmath}
\usepackage{hyperref}
\usepackage{cancel}
\newcommand{\be}{\begin{equation}}
\newcommand{\ee}{\end{equation}}
\newcommand{\bea}{\begin{eqnarray}}
\newcommand{\eea}{\end{eqnarray}}

\newcommand{\dd}{\text{d}}

\begin{document}
\sloppy  

\preprint{LPT--Orsay 18-95, IFT-UAM/CSIC-19-8}

\vspace*{1mm}

\title{A Dark Matter Interpretation of the ANITA Anomalous Events}
\author{Lucien Heurtier$^{a}$}
\email{heurtier@email.arizona.edu}
\author{Yann Mambrini$^{b}$}
\email{yann.mambrini@th.u-psud.fr}
\author{Mathias Pierre$^{b,c,d}$}
\email{mathias.pierre@uam.es}
\vspace{0.5cm}
\affiliation{
${}^a$ 
Department of Physics, University of Arizona, Tucson, AZ   85721}
\affiliation{
${}^b$ 
Laboratoire de Physique Théorique (UMR8627), CNRS, Univ. Paris-Sud, Université Paris-Saclay, 91405 Orsay, France}
\affiliation{
$^c$
Instituto de F\'{i}sica Te\'{o}rica (IFT) UAM-CSIC, Campus de Cantoblanco, 28049 Madrid, Spain} 
\affiliation{
$^d$
Departamento de F\'{i}sica Te\'{o}rica, Universidad Autónoma de Madrid (UAM), Campus de
Cantoblanco, 28049 Madrid, Spain}

\begin{abstract} 
The ANITA collaboration recently reported the detection of two anomalous upward-propagating extensive air showers exiting the Earth with relatively large emergence angles and energies in the range $\mathcal{O}(0.5\!-\!1)~\mathrm{EeV}$. We interpret these two events as coming from the decay of a massive dark-matter candidate ($m_\text{DM}\!\gtrsim\! 10^{9}~\mathrm{GeV}$) decaying into a pair of right-handed neutrinos. While propagating through the Earth, these extremely boosted decay products convert eventually to $\tau$-leptons which lose energy during their propagation and  produce showers in the atmosphere detectable by ANITA at emergence angles larger than what Standard-Model neutrinos could ever produce. We performed Monte Carlo simulations to estimate the propagation and energy loss effects and derived differential effective areas and number of events for the ANITA and the IceCube detectors. Interestingly, the expected number of events for IceCube is of the very same order of magnitude than the number of events observed by ANITA but at larger emergence angles, and energies $\lesssim 0.1~\mathrm{EeV}$. Such features match perfectly with the presence of the two upward-going events IceCube-140109 and IceCube-121205 that have been exhibited from a recent re-analysis of IceCube data samples. 
\end{abstract}

\maketitle

\maketitle

\setcounter{equation}{0}



\section{Introduction}

Despite very indirect but clear astrophysical and cosmological evidences of the presence of dark matter in our Universe, its nature is still unknown~\cite{planck}. Not only the recent direct detection limits obtained by the collaborations LUX~\cite{LUX}, XENON1T~\cite{XENON} and PANDAX~\cite{PANDAX} exclude the simplest electroweak extensions of the Standard Model like Higgs portal~\cite{hp,Higgsportal}, $Z$-portal~\cite{Zportal} or $Z'-$portal \cite{Zpportal}\footnote{see~\cite{review} for a recent review on the subject}but even question the WIMP paradigm. Recently, refined cosmological analysis  on the reheating mechanism~\cite{Garcia:2017tuj} combined with the freeze-in mechanism~\cite{fimp,Bernal:2017kxu} showed that EeV dark-matter particles can be produced from the thermal bath at very early stages of the reheating epoch. From GUT constructions to high-scale supergravity, several motivated models enter in this category (see~\cite{UVfimp,sugra} for some examples). 
Such massive candidates are clearly out of the sensitivity reach of the next generation of direct or indirect detection instruments or at the LHC. Therefore the only clear signatures 
would be through its decay products if the dark matter is unstable.

On the other side, ANITA collaboration confirmed recently the observation of 2 upgoing events, corresponding to energies at the EeV scale~\cite{Gorham:2016zah,Gorham:2018ydl}. A Standard-Model (SM) explanation of these events, considering a SM neutrino of energy $\sim 10^{9}$ GeV crossing the Earth is quite inconceivable since the rock and ice are opaque to the propagation of SM neutrinos at these energies due to the strength of weak interactions. As a matter of fact, at such  energies, the transmission probability $p$ of a SM neutrino through the Earth, on chord length of order 5000-7000 km (corresponding to the arrival directions of these events) is of order $p\simeq \mathcal O(10^{-8}-10^{-6})$.  

Different attempts to explain these EeV events appeared recently. The authors of~\cite{Cherry:2018rxj} (and later on \cite{Nieuwenhuizen:2018vzf}) for instance proposed that sterile neutrinos $\nu_4$ originated by Ultra High Energy (UHE) cosmic rays could propagate inside the earth on a long distance, producing a $\tau$ cascade through charge current interactions with nucleons near the eerth surface. Similarly, in~\cite{Huang:2018als} the author proposed the production of keV right-handed neutrino generated also by UHE cosmic ray to fit the ANITA excesses. In Ref.~\cite{Yin:2018yjn} it was shown that EeV neutrinos could up-scatter on the C$\nu$B into some highly boosted particle which could be long-lived enough to reach the Earth and produce events on Earth measurable by ANITA.
Alternatively,~\cite{Anchordoqui:2018ucj} proposed a decaying superheavy EeV right-handed neutrino trapped in the earth and decaying into active neutrino near the surface and a more exotic sphaleron or leptoquark interpretations were given in~\cite{Anchordoqui:2018ssd,Chauhan:2018lnq} respectively. Some supersymmetric interpretations can be found in the interesting work of \cite{Fox:2018syq} where long-lived right handed staus $\tilde \tau_R$ produced via UHE cosmic rays reach the earth and propagate  before decaying into a gravitino and a $\tau$. Another supersymmetric framework has been proposed in~\cite{Collins:2018jpg} where EeV active neutrinos produce resonantly in the rock long-lived binos (plus sleptons) which in turn decay into a neutrinos near the surface through ${R_p}$-violating couplings.
The only dark matter explanation of the ANITA anomalous events was suggested in~\cite{Anchordoqui:2018ucj}, originating from an atypical dark matter density distribution in the Earth. However, a quantitative analysis of the expected number of events was not explored. The aim of our work
is to provide a detailed quantitative analysis of these EeV anomalous events as potential signatures of decaying dark matter particles, present in the galactic halo, into sterile neutrinos mutating into active ones through their passages in the earth.

The paper is organized as follows. After a summary of the ANITA anomalous events in section I, we introduce our model in section II and compute the galactic production of right-handed neutrino from dark matter decay. We discuss the relevant cosmological constraints in section III before developing the Monte Carlo analysis used to simulate the observed ANITA events in section IV. We then analyze the anomalous events and discuss the detection complementarity with IceCube in section V and VI before concluding.

\vskip 1.5cm

\section{I. The Signal}

The Antarctic Impulsive Transient Antenna (ANITA) is a balloon experiment operating around the south pole, aiming to detect ultra-high energy (UHE) neutrinos by searching for radio pulses produced during their propagation through the Antartic ice. The collaboration has been launching three different flights, for a total duration of $\sim$~85.5 days (see Tab.~\ref{tab:flightsduration}) during which they could collect more than $\gtrsim 30$ events \cite{Gorham:2016zah,Gorham:2018ydl,Hoover:2010qt}. Although most of these events featured a phase reversal characteristic of Extensive Air Showers (EAS) reflecting off the Antartic ice from downward-propagating cosmic rays, the collaboration reported two anomalous $\sim 0.6$ EeV upward-moving events in ANITA I \cite{Gorham:2016zah} and ANITA III \cite{Gorham:2018ydl} flights which do not feature this phase reversal and can therefore be considered as produced by originally upward-propagating cosmic rays\footnote{ ANITA II was not configured to be sensitive to such events.}. More interestingly, these two events could not be associated with any astrophysical point source such as SNe neutrino bursts. Therefore such events are likely to originate from some unknown cosmic-ray neutrino flux. The fourth flight of ANITA has been already realized, and partial analysis of the datas have been published concerning the in-ice Askaryan emission~\cite{Gorham:2019guw} but hasn't reported any results concerning upward-going events yet.

Last but not least, such events have been shown to reach the detector with angles of $-27.4\pm 0.3^\circ$ and $-35.0\pm 0.3^\circ$ under the horizontal\footnote{corresponding to $117^\circ$ and $125^\circ$ zenith angles, and $25.4^\circ$ and $35.5^\circ$ emergence angles, respectively.}, corresponding to showers which escaped the surface of the Earth with emergence angles $25.4^\circ$ and $35.5^\circ$ with a $\sim 1^\circ$ uncertainty \cite{Romero-Wolf:2018zxt}.
Such large zenith angles, corresponding to a passage through the earth of more than 5000 kms of rocks are very challenging in the framework of Standard Model interactions. Indeed, the probability $p$ to observe an EeV $\tau$ emerging through multiple $\nu_\tau-\tau$ regenerations is 
$p \lesssim 10^{-6}$ \cite{Gorham:2018ydl}. To have an idea, the $\nu_\tau$ flux necessary to observe 2 events emerging below ANITA is 12 millions of $\nu_\tau$ per $\mathrm{km^{2}sr^{-1}year^{-1}}$ which is more than 1 million times above the current limit given by Icecube. The authors of \cite{Fox:2018syq} concluded that a Standard Model interpretation of the up-going ANITA events is excluded at the $5\sigma$ level.
\begin{table}[!]
    \centering
    \begin{tabular}{|l|c|c|c|c|}
  \hline
 Flight &ANITA I & ANITA II & ANITA III & ANITA IV \\
  \hline
Duration & 35 days & 28.5 days & 22 days & 29 days \\
\hline
 Events & \#398526 & - & \#1571714 & TBA\\
  \hline
  Energy & $0.6\pm 0.4\mathrm{EeV}$& - & $0.56^{+0.3}_{-0.2}\mathrm{EeV}$&TBA\\
  \hline
  $\theta_{\text{em}}$& $25.4\pm 1^\circ$ &-&$35.5\pm1^\circ $ &TBA\\
  \hline
\end{tabular}
    \caption{Duration of the three flights realized by ANITA for a total of 85.5 operating days.
}
    \label{tab:flightsduration}
\end{table}
\section{II. Galactic production of right handed neutrinos}

\subsection{Decaying Dark Matter hypothesis}

The minimal and simplest extension one can imagine involving a dark-matter scalar field $\phi$ of mass $m_\text{DM}$ and a right-handed neutrino $\nu_R$, both assumed to be SM singlets, can be written\footnote{ We give the example of a real scalar dark matter, but considering a complex scalar would not affect our results.}

\begin{equation}
  {\cal L}= {\cal L}_\text{SM} + {\cal L}_\nu +  
\frac{y_\phi}{\sqrt{2}}\phi \bar\nu_R^c \nu_R - \frac{1}{2} m_\text{DM}^2 \phi^2\,,
\end{equation}
with
\begin{equation}
{\cal L}_\nu= - \frac{1}{2}m_R \bar \nu_R^c \nu_R 
- y_\nu \Bar{L}_L \tilde{H} \nu_R +~\text{h.c.}\,,
\end{equation}
where $y_\nu$ is a Yukawa coupling, $L_L$ denotes a SM leptonic $SU(2)_L$ doublet\footnote{We consider only terms involving one generation of leptons for simplicity.} and  $\tilde{H}$ is the conjugate $SU(2)_L$ Higgs doublet $\tilde{H} \equiv i \sigma^2 H^*$. We  introduced explicitely a Majorana mass  $m_R$ for the right-handed neutrino, allowed by gauge invariance and a Yukawa coupling $y_\phi$. As allowed by gauge invariance, renormalizable operators such as $|H|^2\phi^2$ or $\phi^n$ with $n=1,3,4$ should be considered. However, the detailed study of the scalar potential is beyond the scope of the phenomenological purposes of this paper and would depend on the underlying extended symmetry of a more complete setup. For simplicity, in the following we will just assume that the dark matter field does not acquire a vacuum expectation value.
A similar construction was proposed to interpret high energies events observed by Icecube~\cite{Dudas:2014bca}. 
It is important to notice that even if it seems natural in high-scale versions of the see-saw mechanisms (like $SO(10)$
inspired models for instance) to expect $m_R$ in the range $10^6-10^{10}$ GeV, there are no real experimental nor theoretical constraints on the right-handed neutrino mass which can be as light as $m_R \simeq 1$ eV~\cite{deGouvea:2005er}, implying a sufficiently long-lived right-handed neutrino to cross galactic scales before reaching the earth. 
After diagonalization, the physical states $\nu_1$ and $\nu_2$ of masses $m_1 < m_2$ are defined as~\cite{Dudas:2014bca}
\bea
&&
\nu_1 \simeq \nu_L + \nu_L^c - \theta_R(\nu_R + \nu_R^c) \,,
\nonumber
\\
&&
\nu_2 \simeq \nu_R + \nu_R^c + \theta_R(\nu_L + \nu_L^c)\,,
\nonumber
\eea
\noindent
at leading order in $\theta_R \simeq \sin \theta_R \simeq y_\nu v/m_R$, $v$ being the Higgs vacuum expectation value 
and $\nu_1$ is the Standard Model neutrino. 
The mixing angle $\theta_R$ being experimentally constraints by Planck data to lie in the range
$\theta_R \lesssim 10^{-2}$ for $m_R \gtrsim 1$ eV~\cite{Bridle:2016isd}, 
$\nu_2$ is almost exclusively composed of its right-handed component. From now on we will then denote the physical right-handed neutrino state $N_R \equiv \nu_2$ and the  Standard Model neutrino $\nu \equiv \nu_1$. For simplicity we will consider in our analysis mixing with the $\tau$-neutrino, generalization to three families being straightforward\footnote{In general, mixings with $\nu_e$ and $\nu_\mu$ are also possible. However, such a detailed analysis lies beyond the simplest setup required to explain the ANITA events and is beyond the scope of this work.}.
Moreover, independently on the specific underlying particle physics model, noticing that the most important contribution to the dark-matter decay width is the channel $\Gamma_\textrm{DM}\simeq \Gamma_{\phi \rightarrow N_R N_R}$, one expects the decay channels $\phi \rightarrow N_R \nu$ and $\phi \rightarrow \nu \nu$, to be suppressed respectively by factor of $\theta_{R}^2$ and $\theta_{R}^4$ rendering the condition on the lifetime of the dark matter compatible with indirect constraints from other neutrino observations.

\subsection{The right-handed neutrino induced flux}

\begin{figure}
    \centering
    \includegraphics[width=\linewidth]{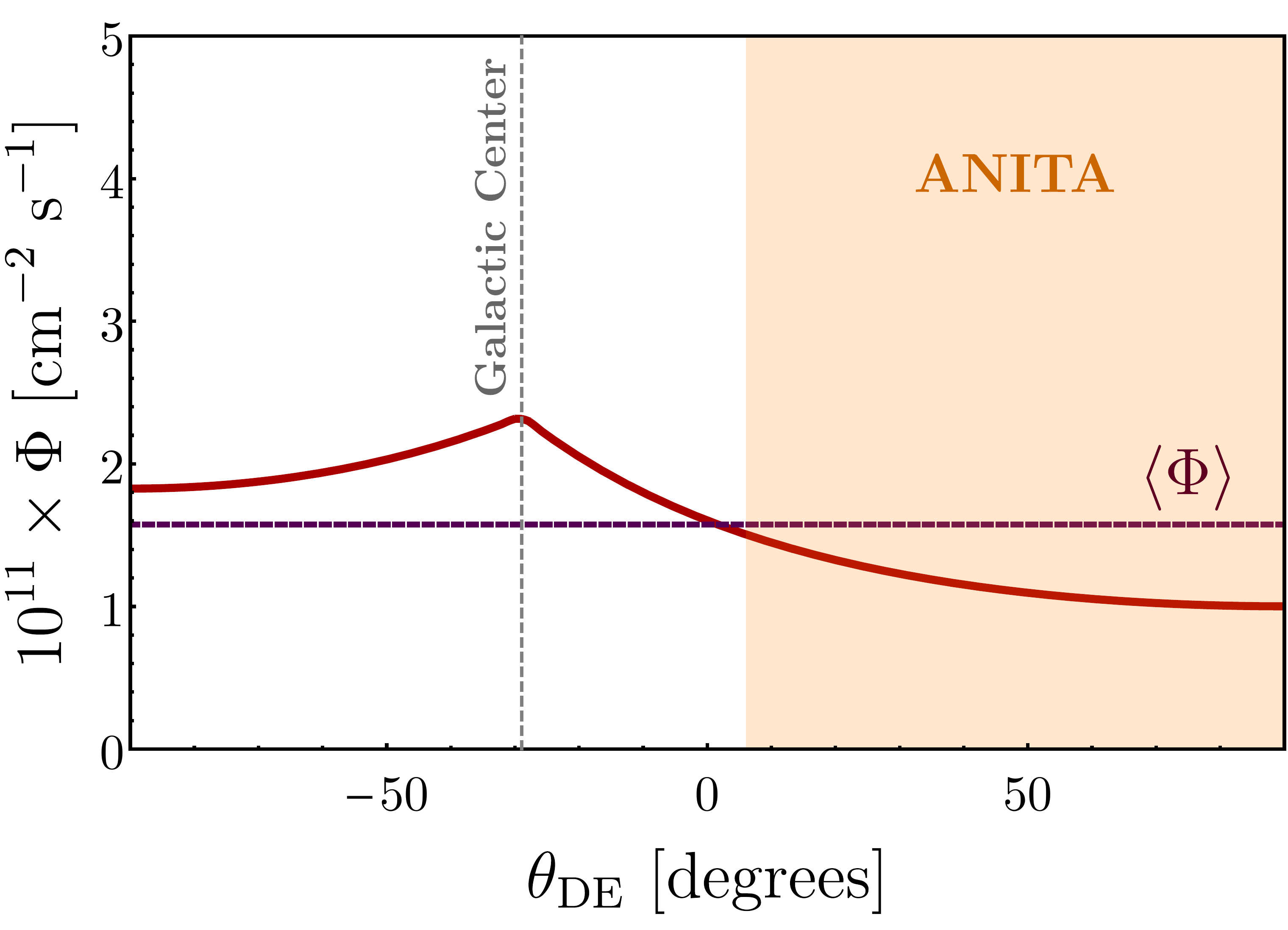}
    \caption{Expected flux of right-handed neutrino on earth, as function of the declination angle $\theta_{\mathrm{DE}}$
    for a $10~\text{EeV}$ dark-matter candidate with a lifetime $\tau_\textrm{DM}=10^{23}$ seconds and a NFW halo profile. The mean flux $\langle \Phi \rangle$, averaged over $\theta_\text{DE},$ is represented in dashed purple. The orange part dubbed "ANITA" corresponds to the region for which events coming from the earth are observable by ANITA.}
  \label{fig:flux}
\end{figure}

The right-handed neutrino flux produced by dark-matter decay reaches the earth surface
with a given direction in the International Celestial Reference System (ICRS) parametrized by the declination ($\theta_\text{DE}$) and the right-ascension ($\phi_\text{RA}$) angles given by

\begin{equation}\label{eq:flux}
    \mathcal{F}(\theta_\text{DE},\phi_\text{RA})=\dfrac{2}{4\pi \tau_\textrm{DM} m_\textrm{DM}}\int_\text{los} \rho_\textrm{DM} \left[r(l,\theta_\text{DE},\phi_\text{RA})\right]~\text{d} \ell~,
\end{equation}

\noindent
where $\tau_\textrm{DM}=\Gamma_\textrm{DM}^{-1}$ is the dark-matter lifetime and $\rho_\textrm{DM}(r)$ is the dark-matter density distribution of the Milky Way, whose radial coordinate is denoted by $r(\ell,\theta_\text{DE},\phi_\text{RA})$\footnote{The coordinates of the Galactic Center (GC) are $r^\text{GC}\simeq8.33~\text{kpc}, \theta_\text{DE}^\text{GC} \simeq -28.92 \degree,\phi_\text{RA}^\text{GC}\simeq 17.86 \degree$ at J2000, see Fig.(\ref{fig:Earth})}. $l$ is a coordinate that runs along the line of sight (los). For our purpose, we parametrize the dark-matter distribution in the Milky Way by a Navarro-Frenk-White (NFW) profile~\cite{Navarro:1995iw} :

\begin{equation}
    \rho_\textrm{DM}(r)\propto \dfrac{1}{\left(\dfrac{r}{r_s} \right) \left[ 1 + \left(\dfrac{r}{r_s} \right)^2 \right]}~,
\end{equation}

\noindent
with $r_s=24~\text{kpc}$ and the distribution is normalized such that the dark-matter density in the vicinity of the solar system is $\rho_\odot=0.3~\text{GeV cm}^{-3}$ \cite{Catena:2009mf} . Dealing with events generated in directions opposite to the galactic center, we checked that the dependence on the profile is completely negligible in our analysis. Since the ANITA experiment is located at the south pole on the Earth rotation axis which is a time-invariant in the ICRS, the coordinate $\phi_\text{RA}$ of any fixed direction in the galaxy varies from 0 to $2\pi$ in 24 hours while the $\theta_\text{DE}$ coordinate remains unchanged. Therefore to deal with a galactic flux, it is more convenient to define a quantity integrated over $\phi_\text{RA}$\footnote{Our approach is valid on average here as the ANITA observation time $~85.5~\text{days}$ is much larger than $24$ hours.} :
\begin{equation}
    \Phi(\theta_\text{DE})=\int \mathcal{F}(\theta_\text{DE},\phi_\text{RA})~\text{d}\phi_\text{RA}~.
\end{equation}
The evolution of this integrated flux with the declination is shown in Fig.~\ref{fig:flux} where $\langle \Phi \rangle$ is the mean flux, averaged over $\theta_{\text{DE}}$ that can be estimated as
\begin{equation}
\langle \Phi \rangle \simeq 1.6 \times 10^{-11}~\text{cm}^{-2}~\text{s}^{-1} \left( \dfrac{10^{23}~\text{s}}{\tau_\textrm{DM}} \right)\left( \dfrac{ 10~\text{EeV}}{m_\textrm{DM}} \right)\,.
\end{equation}

\noindent
For illustration, we plotted the expected right-handed neutrino flux on earth as a function of the declination angle\footnote{In such notations, the horizontal plane for ANITA corresponds to the plane defined by $\theta_\text{DE}=0$. Note that, due to the elevation of ANITA above the Earth of about 35 km, the horizontal ($\theta_{\text{DE}}=0$) does not correspond to the Earth horizon as seen by ANITA, corresponding to a declination angle of $\theta_\text{DE}\approx 6^\circ$} below the horizontal. $\theta_{\text{DE}}$, for a 10 EeV dark-matter
candidate with a lifetime $\tau_{\text{DM}}=10^{23}$ seconds.  Notice that the expected flux is almost isotropic after averaging over $\phi_\text{RA}$, which makes the particular angles favoured by the anomalous ANITA events unlikely to be related to the inhomogeneous dark-matter distribution of the Milky Way\footnote{The consideration of a different galactic DM  profile is not expected to have a significant impact on our results, whereas a different normalization of $\rho_\odot$ would correspond an overall rescaling of the flux.}. It is surprising to observe that a flux of the order of $\langle \Phi \rangle \simeq 1.6 \times 10^{-11}~\mathrm{cm^{-2} s^{-1}}$
corresponds roughly to 2 events in 100 days of observations on a 1 $\mathrm{m}^2$ surface, which is the order of magnitude of the number of events observed by ANITA during its 85.5 days of flights. 

\subsection{Right-handed neutrino propagation in the Galaxy}
The mixing between right-handed neutrino $\nu_R$ and the active neutrino  
 $\nu_L$ renders $N_R$  unstable and allows it to decay into Standard Model lighter states. In order for $N_R$ to propagate through the galaxy and reach the Earth, it is necessary for its decay length to be sufficiently large, typically larger than the size of the dark-matter halo of the Milky Way $\sim 50~\text{kpc}$. As shown further, such constraints restricts the mass  $m_{R}\lesssim 1~\mathrm{GeV}$. For such masses, the $N_R$ decays dominantly into three neutrinos with a decay width

\bea
  \Gamma_{N_R\rightarrow3\nu} &=&\dfrac{G_F^2 m_{R}^5}{32\pi^3} \theta_{R}^2\,, \nonumber\\
  &\simeq& 10^{-22}\left( \dfrac{\theta_{R}}{10^{-2}}\right)^2 \left( \dfrac{m_{R}}{0.1~\text{GeV}}\right)^5~ \mathrm{GeV}\,,
  \label{Eq:gammanu}
\eea

\noindent
and into $N_R \rightarrow \pi^0 \nu, \rho \nu, \nu e^+ e^-$~\cite{Gorbunov:2007ak}. In the Earth frame, the decay length $\lambda$ of a boosted right-handed neutrino with energy $E_{N_R}=m_\text{DM}/2$ produced by the decay of a dark-matter particle at rest is therefore

\begin{equation}
    \lambda\simeq \dfrac{c \gamma}{\Gamma_{N_R \rightarrow3\nu}}\simeq 40~\mathrm{kpc}  \left(\dfrac{10^{-2}}{\theta_R}\right)^{2} \left( \dfrac{22~ \mathrm{MeV}}{m_R}\right)^{6}
    \left(\dfrac{m_\text{DM}}{20~\mathrm{EeV}}\right)
    \,,
    \label{Eq:lambda}
\end{equation}
\noindent
where we introduced the boost factor $\gamma= E_{N_R}/m_{R}$. Requiring the right-handed neutrino to be able to propagate over 50 kpc gives the following condition

\begin{equation}
m_{R} \lesssim 20~\mathrm{MeV} \left( \dfrac{10^{-2}}{\theta_R}\right)^{1/3}  \left( \dfrac{m_\text{DM}}{20~\text{EeV}}\right)^{1/6}.
\label{eq:50kpc}
\end{equation}
A quick look at equations (\ref{Eq:gammanu}) and 
(\ref{eq:50kpc}) helps to understand some tensions that
can appear once we want to interpret the ANITA events with a dark-matter source. Indeed, to generate an event observable by ANITA, $N_R$ should
convert while crossing the earth into a Standard Model neutrino $\nu$.
For that reason, a reasonable mixing angle $\theta_R$ is needed.
On the other hand, a too large mixing angle shorten the lifetime of the right-handed neutrino dangerously, 
reducing drastically its flux on earth.
One then needs a relatively light $N_R$ to compensate the 
effect of $\theta_R$ on its lifetime which restricts the possible mass range to be $m_{R} < 200~\text{MeV}$ for reasonable values of the mixing angle $\theta_{R}\gtrsim 10^{-5}$ and $m_\text{DM}\gtrsim \text{EeV}$, conditions which are required to explain the observed ANITA events as we will describe later.

\section{III. Constraints}

\begin{figure}[t!]
\centering
\includegraphics[width=1\linewidth]{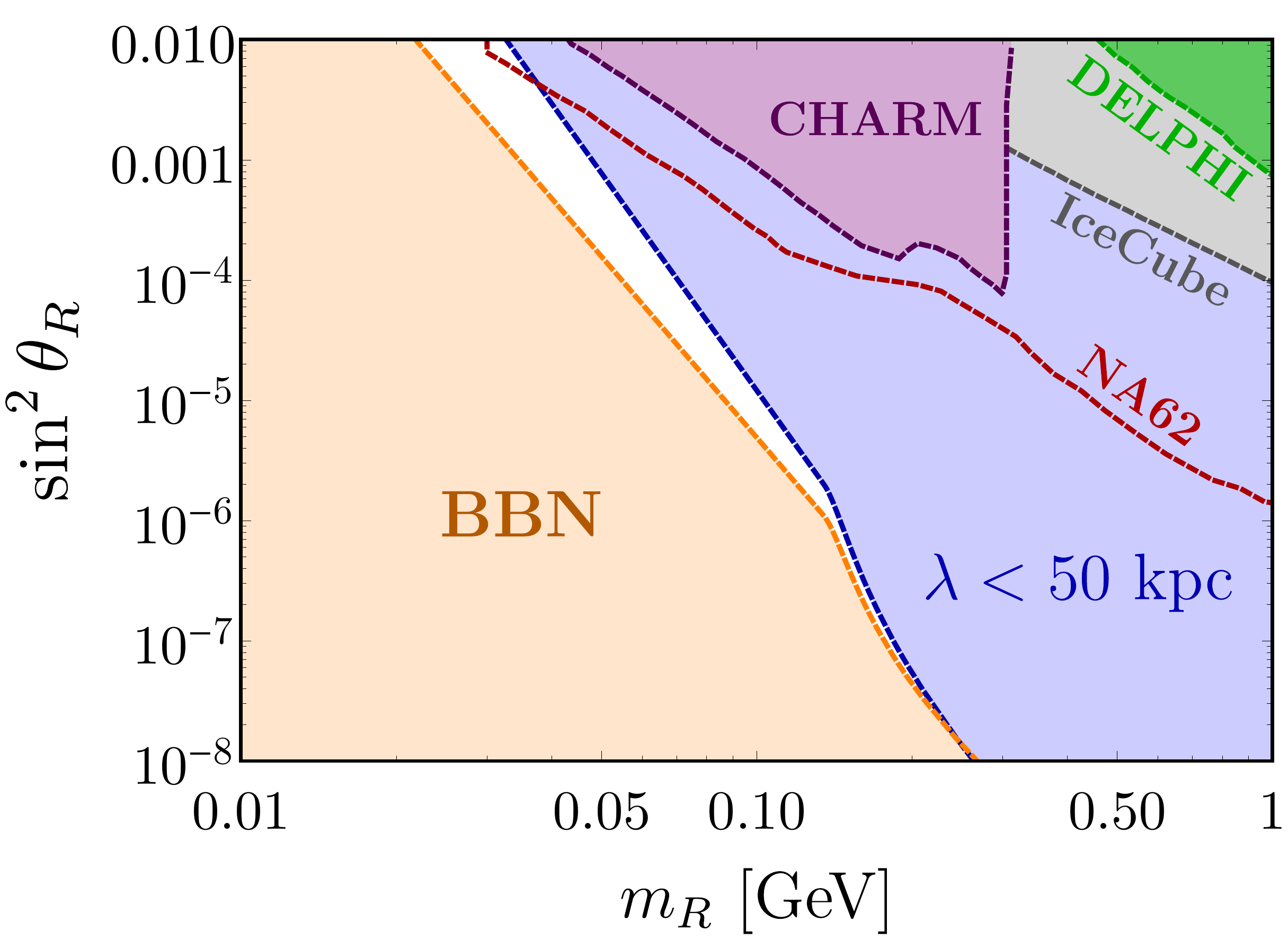}
\caption{\label{fig:BBN}Constraints on the RHN mixing angle for $m_\textrm{DM}=5\times10^4~\text{EeV}$, imposing a galactic propagation length of at least $50$ kpc (in blue), a lifetime shorter respecting BBN constraints~(\ref{eq:BBN}). Constraints from CHARM~\cite{Bergsma:1985is}, DELPHI~\cite{Abreu:1996pa} and IceCube~\cite{Coloma:2017ppo} are represented in purple, green and gray as well as predictions for the NA62 experiments~\cite{Drewes:2018gkc} represented in red.}
\end{figure}

The introduction of light species with  sizable couplings to SM particles could have dramatic consequences on the cosmological history of the Universe. Indeed, the presence of a Right-Handed Neutrino (RHN) population might cause a substantial increase of the value of the effective number of relativistic species $N_\text{eff}$, resulting in the acceleration of the universe expansion, affecting the predictions of light element abundances from the Big Bang Nucleosynthesis (BBN), the Baryon Acoustic Oscillation (BAO) scale or the present value of the Hubble constant $H_0$~\cite{Ruchayskiy:2012si,Vincent:2014rja}. In order to be safe from cosmological constraints, we require the RHN to decay before BBN by imposing the conservative condition $\tau_{N_R} \lesssim$ 0.1 seconds. From Eq.~(\ref{Eq:gammanu}),
this corresponds to the bound
\begin{equation}
\tau_{N_R} \lesssim 0.1~\mathrm{seconds} ~~\Rightarrow ~~
\left(\frac{\theta_R}{10^{-2}}\right)^2 
\left( \frac{m_R}{0.1~\mathrm{GeV}} \right)^5 \gtrsim 
0.07\,,
\label{eq:BBN}
\end{equation}
implying $m_R \gtrsim 60$ MeV for $\theta_R = 0.01$.\\
Another possiblity to be safe from cosmological constraints is to consider a less abundant RHN population with a lifetime larger than the age of the universe, which can be achieved for $m_R \lesssim 10~\text{keV}$ and $\theta_R \gtrsim 10^{-5}$. However, a RHN with a mass $m_R \sim ~\text{keV}$ would behave as a cold/warm dark-matter component and would be in conflict with cosmological observations~\cite{Dodelson:1993je,Vincent:2014rja}. Therefore, the only region left for reasonable value of $\theta_R \gtrsim 10^{-5}$ corresponds to the conservative bound $m_R \lesssim 50 $ eV\footnote{For $\theta_R \sim 10^{-4}-10^{-5}$, a RHN with mass $m_R \in [50,100]~\text{keV}$ might be cosmologically viable but would require a dedicated analysis, which is beyond the scope of this paper.}. In this case, the lifetime of the right-handed neutrino is sufficiently large to ensure that it propagates through the Milky Way halo and reaches the earth. Details regarding the two scenarios ($m_R \gtrsim 50$ MeV and $m_R \lesssim 50$ eV) are discussed in the next sections.

\subsection{The $m_R \gtrsim 50 ~\text{MeV}$ window}

In this part of the parameter space, satisfying simultaneously the conditions~(\ref{eq:50kpc}) and~(\ref{eq:BBN}) implies the following condition on the dark-matter mass $m_\text{DM}\gtrsim 10^{4}~\text{EeV}$, which is required to generate a sizable boost factor for the produced right-handed neutrino to travel throughout the galaxy before decaying, whilst evading strong bounds from BBN. In Fig.~\ref{fig:BBN} we represented these constraints for a 50 ZeV dark matter, together with bounds on $\theta_R$ from the CHARM~\cite{Bergsma:1985is} and DELPHI~\cite{Abreu:1996pa} collaborations, as well as double-cascade searches at IceCube~\cite{Coloma:2017ppo} and predictions for the NA62 experiments~\cite{Drewes:2018gkc}, which are less constraining than the requirement $\lambda>50~\text{kpc}$.
Fig.~\ref{fig:BBN} shows that constraints on the parameter space suggest a restrained window for the right-handed neutrino mass $m_R\in[0.02,0.2 ]~\text{GeV}$ while the mixing angle can take any arbitrary value $\theta_R^2>10^{-7}$. Interestingly, the allowed part of the parameter space points towards a region allowing for a successful low-scale leptogenesis in models involving extra heavy neutral states~\cite{Abada:2018oly}.
Such a large dark-matter mass $m_\text{DM}\gtrsim 10^{4}~\text{EeV}$ (and right-handed neutrino energy) could at first sight look far too much for explaining the anomalous events seen by ANITA, but we will see that it actually can perfectly interpret the signal, due to the energy loss that the right-handed neutrino and its decay products undergo while propagating across the Earth crust. For completeness, and as a comparison of our simulation results with previous studies, we will also probe the parameter space at energies $m_\text{DM} \sim 10~\mathrm{EeV}$ and see how the effect of the mixing angle affects the propagation through the Earth in what follows.

\subsection{The $m_R \lesssim 50~\text{eV}$ window}

\begin{figure}[t!]
\begin{center}
\includegraphics[width=1\linewidth]{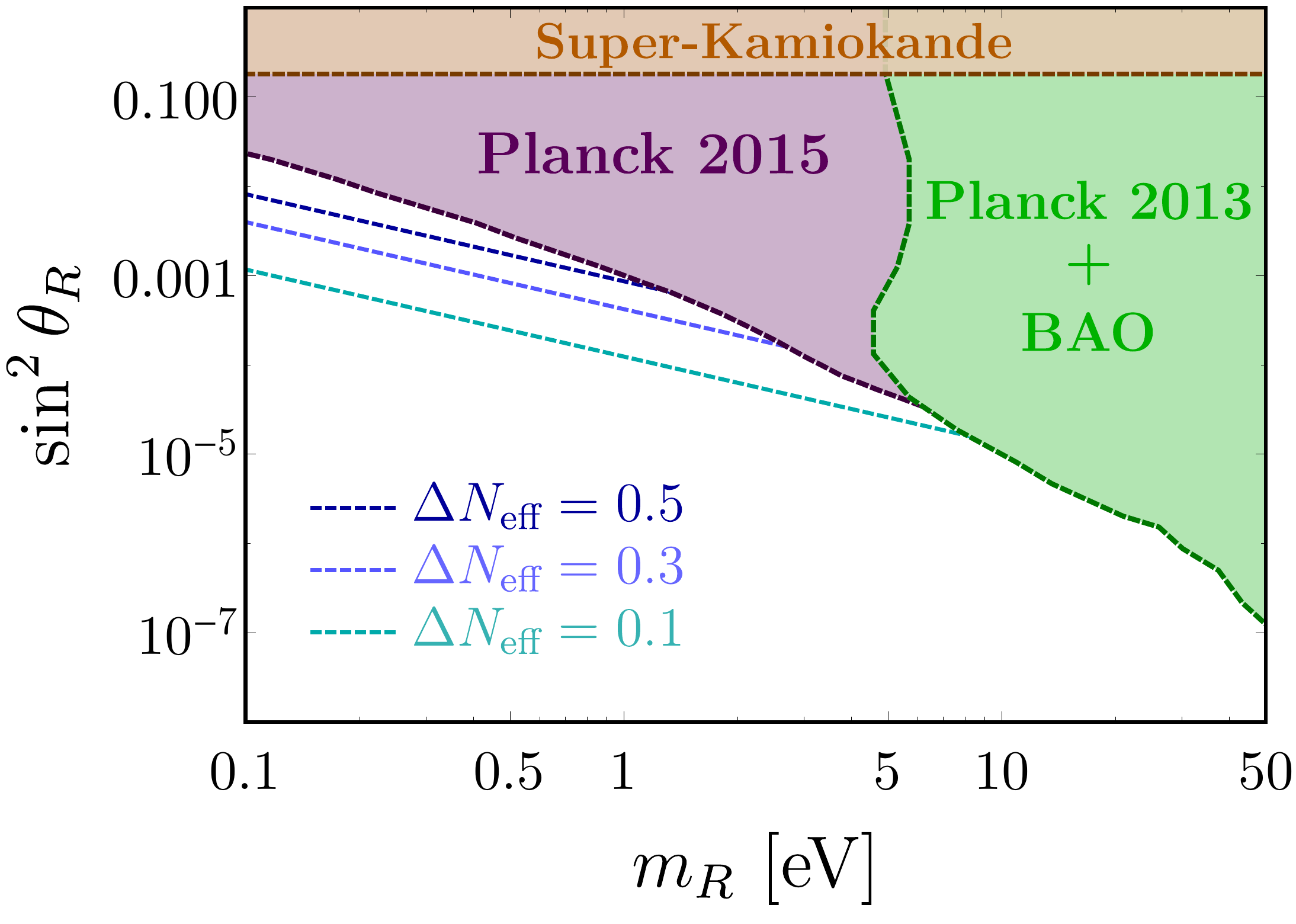}
\caption{\label{fig:ConstraintseVsteriles}Constraints on the RHN  mixing angle and masses from cosmological observations, assuming a Dodelson-Widrow production, and Super-Kamiokande bounds represented in brown~\cite{Abe:2014gda}. Constraints from a CMB analysis based on Planck measurements performed in~\cite{Bridle:2016isd} is represented in purple and dubbed "Planck 2015". Predicted values of $\Delta N_\text{eff}$ from~\cite{Bridle:2016isd} are represented in dashed blue lines. The green region dubbed "Planck 2013 + BAO" correspond to the constraints derived in the analysis~\cite{Vincent:2014rja} including early results from Planck, WMAP9 and BAO constraints.}
\end{center}
\end{figure}

While the mixing angle $\theta_R$ for such light $N_R$ is rather unconstrained from experiments\footnote{As we assumed a mixing mostly with the $\tau$-neutrino}, as discussed further on, the precise value of the mixing angle does not play a significant role on our analysis. The stronger constraint on the mixing angle is derived by the Super-Kamiokande experiment which excludes mixing angles $\sin^2 \theta_R > 0.18$~\cite{Abe:2014gda}. Nevertheless, such light species could have a possible impact on cosmological observables~\cite{Acero:2008rh,Gariazzo:2015rra} through an extra contribution to the effective number of relativistic species $\Delta N_\text{eff}$ or via its energy density at the present time whose effect can be accounted for by considering an extra effective neutrino mass $m_\nu^\text{eff}=\Delta N_\text{eff}~m_R$\footnote{Cosmological constraints are expected to depend on the specific RHN production process, in our case we assume a production via the Dodelson-Widrow mechanism, as expected if no extra state is introduced~\cite{Dodelson:1993je}.}. Such effects are expected to alter the $\Lambda$CDM-expected CMB spectrum by affecting the matter-radiation equality redshift, resulting in an overall shift of the peaks to higher multipoles. Constraints from the analysis performed in~\cite{Vincent:2014rja}, from early results of the Planck CMB measurements, WMAP9 and BAO is represented in Fig.~\ref{fig:ConstraintseVsteriles} which shows as well constraints derived in light of the latest Planck results~\cite{Bridle:2016isd,Knee:2018rvj}. 

One would typically expect extra particles, free-streaming in the universe, to suppress the matter power spectrum on small physical scales and therefore to be constrained by the Lyman-$\alpha$ forest measurements. However, the typically small right-handed neutrino energy density generated by the Dodelson-Widrow mechanism~\cite{Dodelson:1993je} in the allowed parameter of Fig.~\ref{fig:ConstraintseVsteriles} is not sufficient to alter the matter power spectrum in the sensitivity reach of the current Lyman-$\alpha$ measurements~\cite{Baur:2017stq,Acero:2008rh}.

The existence of light species has recently presented some interests in the cosmology community  as a possible solution to a set of discrepancies between several cosmological observations. Indeed, it was shown that a non-negligible value of $\Delta N_\text{eff} \sim 0.2-0.5$ could alleviate the $\sim 4 \sigma$ tensions~\cite{Bernal:2016gxb} between local measurement of the local Hubble constant $H_0$~\cite{Riess:2016jrr,Riess:2018byc} and the value inferred from CMB observations~\cite{Aghanim:2018eyx}. Such tensions have been addressed by introducing extra light degrees of freedom or in extended cosmology framework~\cite{DeltaNeff} but interestingly the prediction of a non-negligible value of $\Delta N_\text{eff}$ appears naturally in some part of our parameter space as represented in Fig.~\ref{fig:ConstraintseVsteriles}.

Moreover, the long-standing discrepancy between large scale structure surveys and the CMB determination of the amplitude of matter density fluctuations in spheres with radius of $8h^{-1}~\text{Mpc}$~\cite{LSSsurveys} as well as a $\sim 2.5\sigma$ tension between the BOSS DR11 BAO measurements from Lyman-$\alpha$ with the $\Lambda$CDM predictions from Planck~\cite{Delubac:2014aqe} have been adressed in~\cite{Poulin:2018zxs}. It was shown that the best fit model for alleviating simultaneously such tensions points towards a sum of neutrino masses of $\sim 0.4~\text{eV}$ which could be effectively achieved within our model.

Another source of interest for eV scale right-handed neutrino concerns recent measurements of neutrino experiments. In particular, the $2-4\sigma$ discrepancy from the so-called short baseline (SBL) neutrino experiments for which a global fit analysis suggests the existence of a sterile neutrino of mass close to the eV scale~\cite{Collin:2016rao}. Recently, IceCube capacities of searching for right-handed neutrino mixing through matter-effects induced mixing to $\tau$-neutrinos have been investigated in~\cite{Blennow:2018hto} for $m_R>10~\text{eV}$ and it was shown that a non-zero mixing is mildly preferred to the non-mixing case and compatible with $m_R\sim 1~\text{eV}$ and a mixing angle $\theta_R\simeq 10^{-1}$ as considered in the ANITA anomalous events interpretation presented in~\cite{Cherry:2018rxj}. However, such a large mixing is in conflict with cosmological observations based on a Dodelson-Widrow production mechanism and would require the presence of supplementary degrees of freedom and a more exotic cosmological history, which is beyond the scope of the minimalist approach considered in this work.\\

\section{IV. Simulating ANITA events}
    
\begin{figure}[t!]
        \centering
        \includegraphics[width=0.8\linewidth]{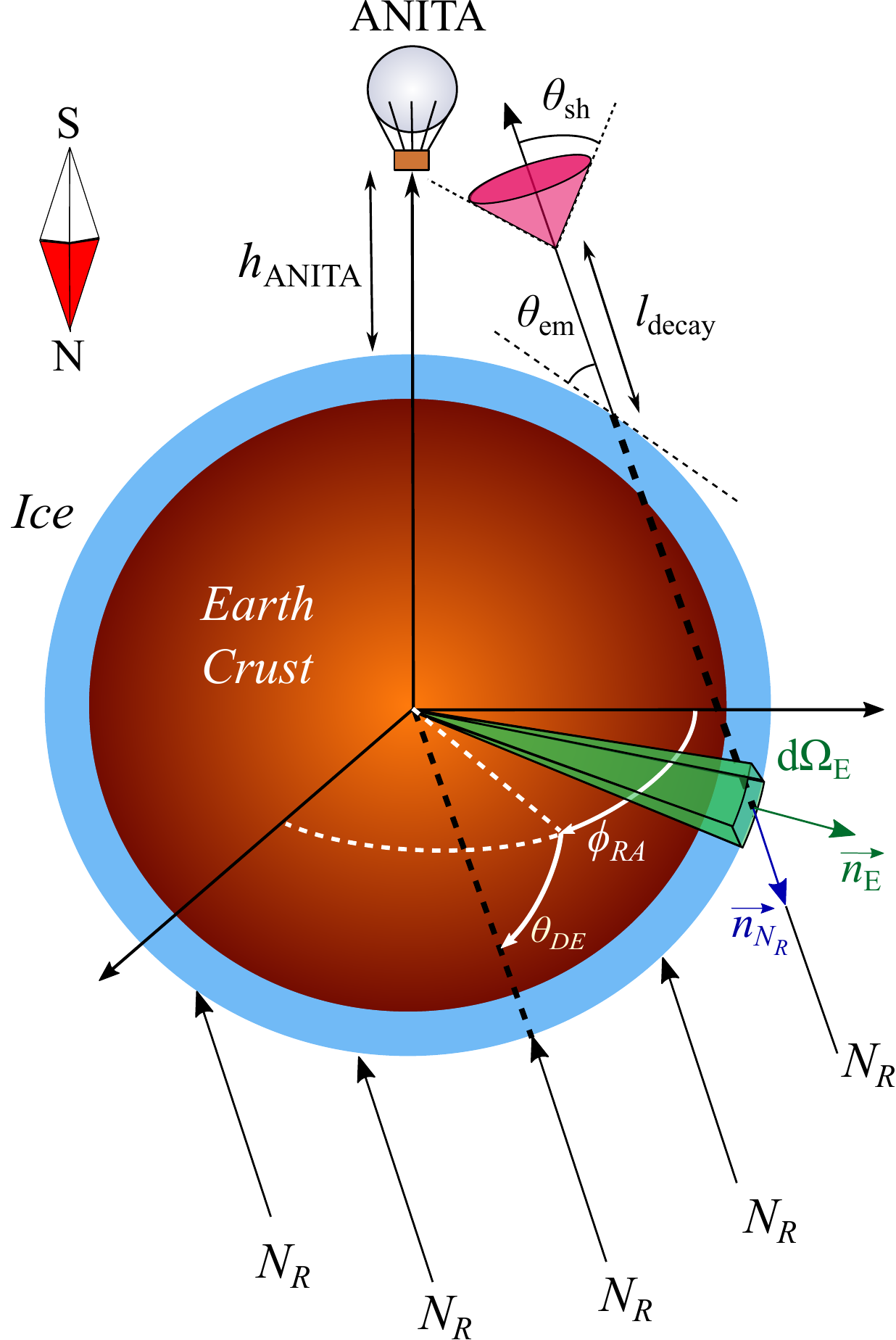}
        \caption{\footnotesize Illustration of the ANITA experiment and useful notations for our simulation understanding.}
        \label{fig:Earth}
\end{figure}

In order to predict the number of events per emergence angle that a dark-matter decay would have produced in the ANITA detector, one needs to properly describe how an $\mathrm{EeV}$ right-handed neutrino could propagate through the Earth crust, scatter into a $\tau$ or $\tau$-neutrino, and produce an Extensive Air Shower (EAS) event in the atmosphere, which could be detected by ANITA $\sim 36$ kilometers above the Earth surface. Such number of events can be calculated using the formula
\begin{widetext}
\begin{equation}
    \frac{\dd N}{\dd\theta_{\text{em}}}=\int \dd E_{\text{exit}} \int \frac{\dd^2 A_{\text{eff}}}{\dd{E}_{\text{exit}}
    \dd\theta_{\text{em}}}({E}_{\text{exit}},\theta_{\text{em}}~|~{E}_{N_R},\theta_{\text{DE}},\phi_{\text{RA}})~\times~ \mathcal{F}(\theta_{\text{DE}},\phi_{\text{RA}})~\times~T_{\text{exp}}\times\sin\theta_{\text{DE}}\ \dd\theta_{\text{DE}} \,\dd\phi_{\text{RA}} 
\end{equation}
\end{widetext}
where $\mathcal{F}(\theta_{\text{DE}},\phi_{\text{RA}})$ is the flux of incoming right-handed neutrinos given in Eq.~\eqref{eq:flux} and the exposure time $T_{\text{exp}}=85.5$~days is the total operating time of the three ANITA flights as calculated from Tab.~\ref{tab:flightsduration}. $\theta_\text{em}$ is the angle of the emerging $\tau$, formed between the direction of propagation and the tangential plane to the Earth at the point where the $\tau$ exits the Earth.
The effective area $A_{\text{eff}}$ denotes the elementary area of the detector for an incoming right-handed neutrino direction  $(\theta_{\text{DE}},\phi_{\text{RA}})$ -- taken in the ICRS coordinates system -- which exits the Earth crust in the form of the $\tau$-lepton of energy ${E}_{\text{exit}}$. Following Ref.~\cite{Romero-Wolf:2018zxt} one can obtain this effective area for every direction of incoming right-handed neutrinos by integrating over the Earth surface the probability that a neutrino entering the Earth can $(i)$ propagate through the Earth and produce a $\tau$-lepton, $(ii)$ the $\tau$ escapes the ice without losing too much energy and $(iii)$ the $\tau$ decays in the atmosphere at a relatively low altitude and $(iv)$ the shower produced by the decay can be detected by ANITA:

\begin{widetext}
    \begin{eqnarray}\label{eq:Aeff}
        \frac{\dd^2 A_{\text{eff}}}{\dd{E}_{\text{exit}}
        \dd\theta_{\text{em}}}({E}_{\text{exit}},\theta_{\text{em}}~|~{E}_{N_R},\theta_{\text{DE}},\phi_{\text{RA}})&=&R_{\text{E}}^2\int \dd\Omega_{\text{E}}~ \vec{n}_{N_R}\cdot\vec{n}_\text{E}
        \times \int \dd{E}_{\text{exit}}\frac{\dd P_\text{exit}}{\dd{E}_{\text{exit}}}({E}_{\text{exit}},\theta_{\text{em}}~|~{E}_{N_R},\theta_{\text{DE}},\phi_{\text{RA}},\theta_{\text{E}},\phi_{\text{E}})\nonumber\\
        &\times&\int \frac{\dd P_{\text{decay}}}{\dd l}(l~|~{E}_{\text{exit}}) \times P_{\text{det}}(\theta_{\text{sh}}|l,\theta_{\text{DE}},\phi_{\text{RA}},\theta_{\text{E}},\phi_{\text{E}})~ \dd l\,,
    \end{eqnarray}
\end{widetext}

\noindent
where $\vec{n}_{N_R}$ and $\vec{n}_\text{E}$ are unitary vectors, exiting the Earth surface from the right-handed neutrino incidence point, $\vec{n}_{N_R}$ is in the direction $(\theta_{\text{DE}},\phi_{\text{RA}})$ and $\vec{n}_\text{E}$ is normal to the Earth surface in the direction $(\theta_\text{E},\phi_\text{E})$. The elementary solid angle corresponding to these angles is denoted by $\dd\Omega_E$. In this expression, the factor $\dd P_\text{exit}/\dd{E}_{\text{exit}}$ encapsulates the probability that an incoming RH neutrino of energy ${E}_{N_{R}}$ propagates through the Earth and converts (after one, or more interactions) into a $\tau$ lepton which manages to escape the Earth after losing a significant amount of Energy during its propagation through the crust. The factor $\dd P_{\text{decay}}/\dd l$ denotes the probability, once the $\tau$ exits the Earth, and that it decays after travelling along a distance $l$ (See Fig.~\ref{fig:Earth}). The probability $P_\text{det}$ finally encodes the capacity for a given shower to produce a sufficient electric field in a direction which would reach the payload and therefore be detected by ANITA.
\subsection{Exit Probability}\label{subsec:dPexit}
\begin{figure*}
    \centering
    \includegraphics[width=0.558\linewidth]{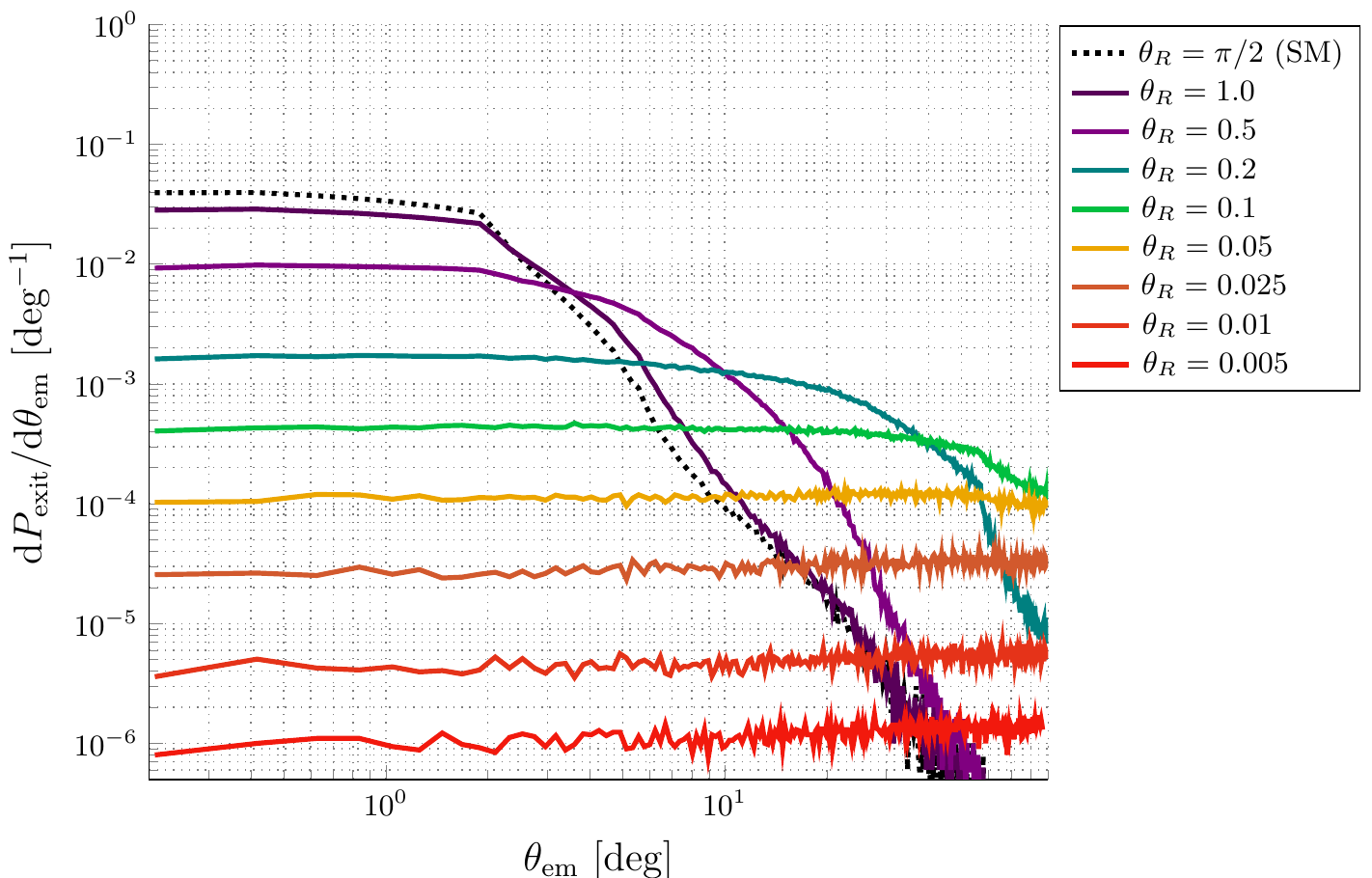}\hspace{0.01\linewidth}\includegraphics[width=0.43\linewidth]{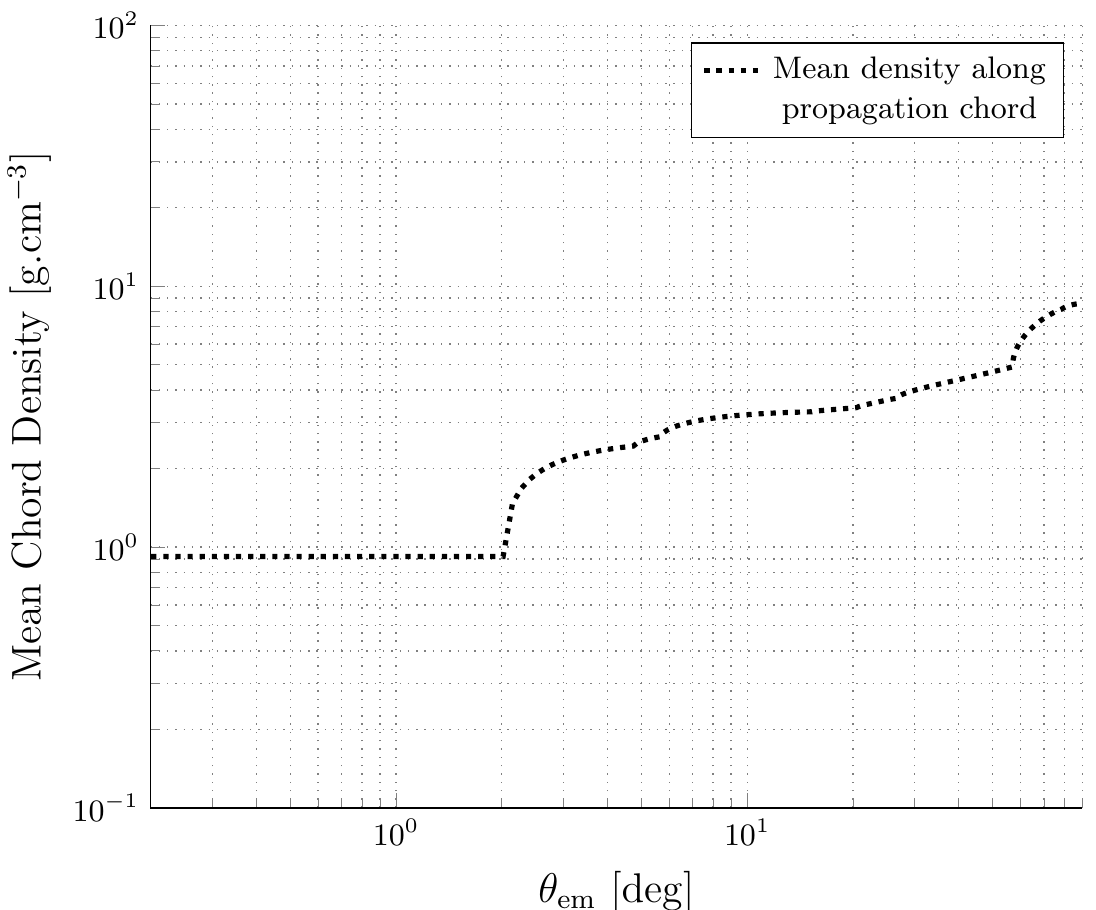}
    \caption{\footnotesize (left panel) Probabilities for an incoming RH neutrino of energy ${E}_{N_{R}}=10~\mathrm{EeV}$ to convert into a $\tau$ and exit the Earth with an energy $>100~\mathrm{TeV}$. (right panel) Mean density crossed by the neutrino during proparation through the Earth as a function of the emergence angle. }
    \label{fig:dPexit}
\end{figure*}
In order to compute the exit probability $\dd P_{\text{exit}}/\dd E_{\text{exit}}$, we made use of a modified version of the publicly available code provided with Ref.~\cite{Alvarez-Muniz:2017mpk} that we adapted to the case of a right-handed neutrino entering the Earth and converting into a $\tau$-neutrino or $\tau$-lepton through the usual neutral current (NC) or charged current (CC) interactions up to a factor $\sin^2 \theta_R$\footnote{In this analysis, for simplicity, we have not considered matter-induced enhancement/suppresion factors of the mixing angle due to the propagation of $N_R$ through the Earth}. The code simulates the probability of transmission through the Earth using a Monte-Carlo approach and assuming an ice layer of thickness at the surface of the Earth. We included in the simulation the possible regeneration of the $\tau$ into a $\nu_{\tau}$ once it decays inside the Earth. However one should note that the regeneration effect starts being negligible when the mixing angle $\theta_R$ is small enough, since the Earth becomes relatively transparent at large emergence angles as compared to the SM case. Therefore the more transparent the Earth is, the less the regeneration of $\tau$-neutrinos plays a significant role in the propagation. 

In Fig.~\ref{fig:dPexit} we present the results obtained for the exit probability $\dd P_{\text{exit}}/\dd{E}_\text{exit}$ for a RH neutrino of energy  ${E}_{N_{R}}=10~\mathrm{EeV}$ as a function of the emergence angle $\theta_{\text{em}}$ and for various mixing angles $\theta_{R}$. One can see that the suppression of the probability $P_{\text{exit}}$ at large emergence angles $\theta_{\text{em}}$ is less and less efficient as compared to small emergence angles while the mixing angle $\theta_{R}$ decreases, since the Earth core becomes transparent enough to let high-energy neutrinos propagate. Inversely, the transmission probability at low emergence angles decreases while the mixing angle decreases, since the low-density layers become too much transparent for the RH neutrino to interact enough in order to produce a lepton before escaping the Earth surface. On the right panel of Fig.~\ref{fig:dPexit} we indicate the mean density crossed by the RH neutrino on its way through the Earth as a function of the emergence angle. The ankle present at $2^\circ$ in the probability distribution in the SM case and for large mixing angles was already pointed out in Ref.~\cite{Alvarez-Muniz:2017mpk} and corresponds to the angle at which the neutrinos start crossing the first rock layer as opposed to neutrinos arriving with low emergence angles $\theta_{\text{em}}\lesssim 2^\circ$ which only propagate through the ice. For lower mixing angles $\theta_R\lesssim 0.5$ one can notice another suppression around $\theta_{\text{em}}\sim 60^\circ$ corresponding to a second significant increase of the mean density along the propagation chord as can be seen in the right panel of Fig.~\ref{fig:dPexit}.
\subsection{Energy Loss during the $\tau$ propagation}
\begin{figure}[t]
    \centering
    \includegraphics[width=1\linewidth]{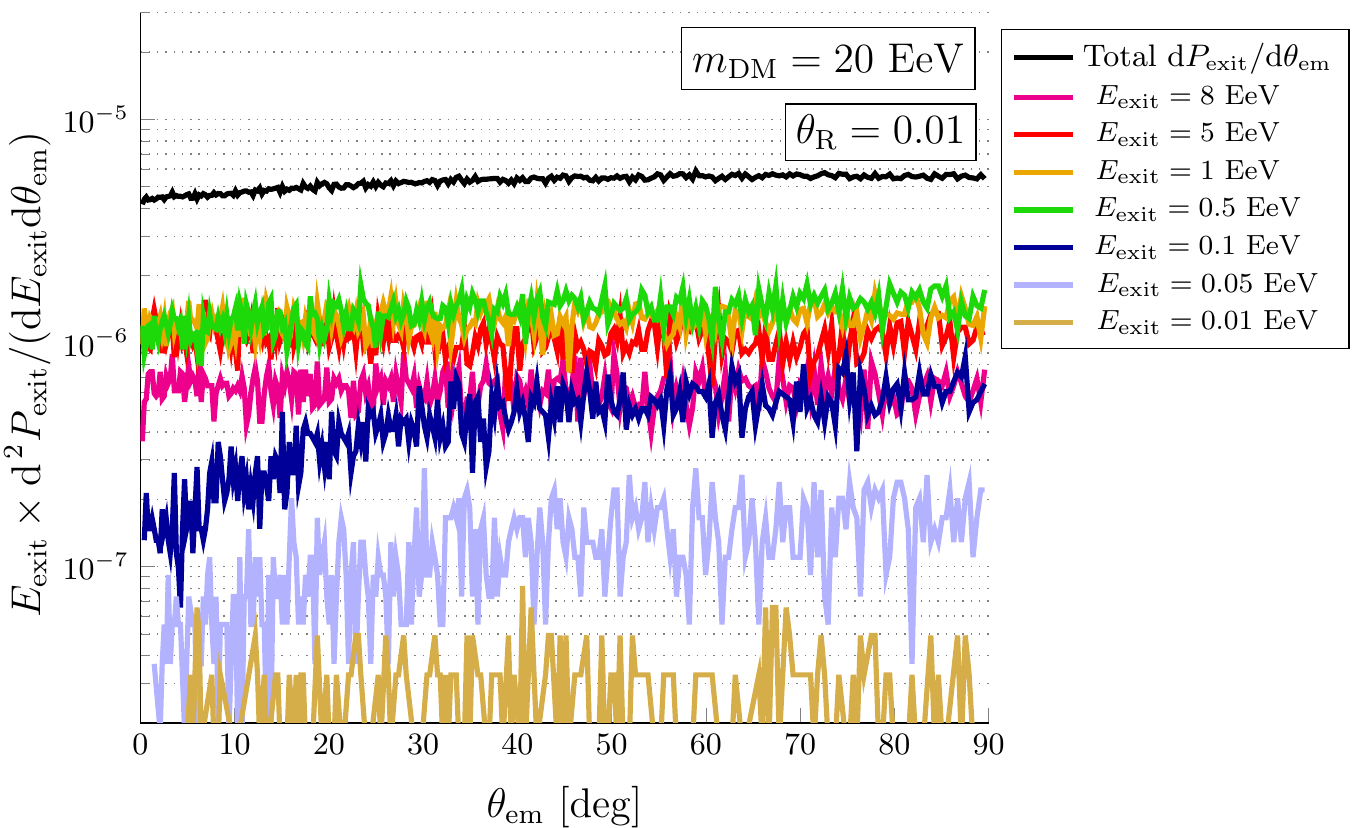}
    \caption{\footnotesize  Exit probability per emergence angle  for given values of the exit energy of a $\tau$-lepton produced by an incoming RH neutrino of energy ${E}_{N_{R}}=10~\mathrm{EeV}$    }
    \label{fig:EnergyLoss}
\end{figure}

As it was already shown in Ref.~\cite{Romero-Wolf:2018zxt}, the deeper the $\tau$-lepton gets produced in the Earth crust, the lower its exit energy ${E}_{\text{exit}}$ is. Therefore, the energy of the events detected by the ANITA experiment can be significantly lower than the energy of the incoming right-handed neutrino produced by dark-matter annihilation. In other words the dark-matter mass could in principle be much larger than the $\mathrm{EeV}$ scale while still generating EAS at an energy range detectable by ANITA. In Fig.~\ref{fig:EnergyLoss} we plot the differential $\tau$-exit probability $\dd ^2P_{\text{exit}}(E_{\text{exit}},\theta_{\text{em}})$ after a right-handed neutrino of energy ${E}_{N_R}=10~\mathrm{EeV}$ crosses the Earth, corresponding in our Monte-Carlo simulation to the fraction of events escaping the Earth with energy  $E\in [E_{\text{exit}}-\frac{\Delta E_{\text{exit}}}{2},E_{\text{exit}}+\frac{\Delta E_{\text{exit}}}{2}]$ where $\Delta E_{\text{exit}} \propto E_{\text{exit}}$ is the step of discretization we chose for scanning over the exit energies. We show the results for the case of a mixing angle $\theta_{R}=0.01$.  As a result, a very few $\tau$'s of energies $\gtrsim 5~\mathrm{EeV}$ manage to escape the crust as compared to leptons whose energies is in the range $0.5-1~\mathrm{EeV}$ which is the region of interest for what concerns the events detected by ANITA. One can moreover notice in Fig.~\ref{fig:EnergyLoss} that below $\lesssim 0.1~\mathrm{EeV}$ the probability of exit starts going down again, which confirms that most of the events that ANITA should observe are expected to be in the range $0.1-1~\mathrm{EeV}$ for an incoming neutrino of $10~\mathrm{EeV}$.

\subsection{Tau Decay in the Atmosphere}    
Once a $\tau$-lepton exits the Earth, its decay probability depends on its energy ${E}_{\text{exit}}$ and varies with the distance on which it has already been propagating without decaying
\begin{equation}
    \frac{\text{d}P_{\text{decay}}}{\text{d}l}(l~|~E_{\text{exit}})=\frac{1}{L(E_{\text{exit}})}\exp\left(-\frac{l}{L(E_{\text{exit}})}\right)\,,
\end{equation}
where the decay length of the $\tau$ in the lab frame is given by $L({E}_{\text{exit}})\sim 49\mathrm{km}\times ( {E}_{\text{exit}}/\mathrm{EeV})$. The energy dependency of the decay length is of crucial importance, since the larger the exit energy is, the more the the exiting $\tau$ is long lived, and the more unlikely it will decay before escaping the atmosphere. As we will see, although the anomalous events observed by ANITA can be interpreted as the decay of a DM particle of mass $\gtrsim 1~\mathrm{EeV}$, such effect will allow us to also interpret the anomalous events of ANITA as produced by $N_R$ of a much larger energies than the EeV scale.

\subsection{Detection Probability}

Once a $\tau$ exits the Earth in a certain direction and decays into an EAS in the atmosphere, the probability that the shower reaches the payload and that the local electric field as measured by ANITA is sufficient for the experimental device to trigger is given by the probability $P_\text{det}(\theta_{\text{sh}}|l,\theta_{\text{DE}},\phi_{\text{RA}},\theta_{\text{E}},\phi_{\text{E}})$. In Ref.~\cite{Romero-Wolf:2018zxt}, a detailed analysis of such probability is presented and led to the conclusion that $(i)$ showers which are produced at altitudes larger than $5-6$ km see their peak electric field decrease since the EAS cannot fully develop in the absence of atmosphere. $(ii)$ The electric-field peak value is spread over an opening angle of about $1^\circ-2^\circ$. $(iii)$ The lower is the energy of the $\tau$ the lower is the electric field measured by ANITA, rendering a triggering of the detector less unlikely for low energetic decaying $\tau$. $(iv)$ Similarly the further away from the detector the shower production is, the more the electric field gets diluted while propagating through the air, and the more unlikely a detection by ANITA is. For simplicity, and since Ref.~\cite{Romero-Wolf:2018zxt} does not provide any analytic estimation of such effect for arbitrary emergence angles, distances of emission from the detector and arbitrary altitude of the decay, we made in what follows the assumption that the showers which are produced at an altitude lower than 15km and for which the location of the ANITA detector is contained in a cone of opening angle $\theta_{\text{sh}}$ with respect to the shower axis of less than $1.5^\circ$ (see Fig.~\ref{fig:Earth} for details) are detected with a probability equal to one. A full EAS shower simulation would be necessary to avoid such simplifying assumption but is beyond the scope of this paper.

\subsection{Effective Area of the Detector}
\begin{figure*}
    \centering
    \includegraphics[width=0.45\linewidth]{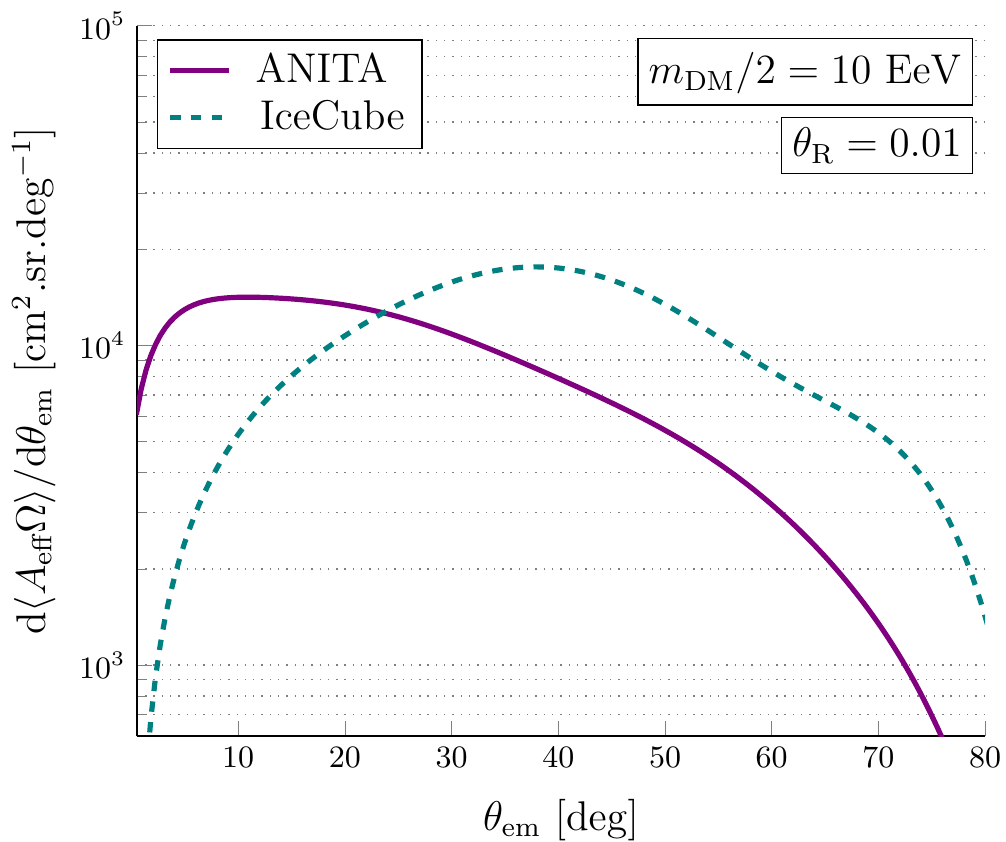}\hspace{0.05\linewidth}\includegraphics[width=0.45\linewidth]{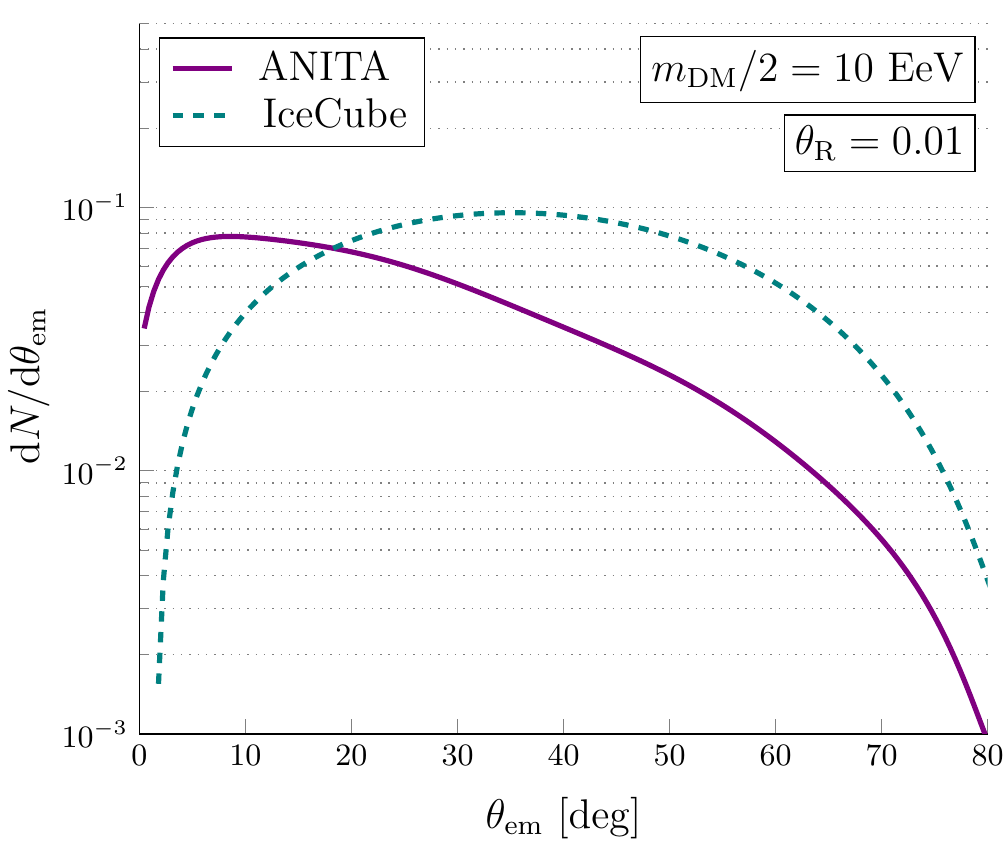}
    \caption{\footnotesize (Left) Effective area of the ANITA and IceCube detectors for a flux of incoming RH neutrinos of energies ${E}_{N_R}=10~\mathrm{EeV}$ and for $\theta_{R}=0.01$. (Right) Number of events per emergence angle expected for IceCube and ANITA after 3142.5 days and 85.5 days of exposure, respectively, and a dark-matter lifetime of $10^{23}~\mathrm{s}$.}
    \label{fig:Aeff}
\end{figure*}

Given the exit probability $\dd P_{\text{exit}}/\dd{E}_\text{exit}$ that we computed earlier in this section, one can obtain the differential effective area of the ANITA detector, as calculated from Eq.~\eqref{eq:Aeff}.  Our results are presented in the left panel of Fig.~\ref{fig:Aeff} in which we show the quantity $\dd \langle A_\text{eff} \Omega \rangle/\dd \theta_\text{em}$ for a given angle $\theta_\text{em}$ where the effective area is integrated over the arrival directions $(\theta_{\text{DE}},\phi_{\text{RA}})$ and over all exit energies $E_{\text{exit}}$ for a mixing angle $\theta_{R}=0.01$. Interestingly enough, when decreasing the mixing angle to small enough values, as illustrated in Fig.~\ref{fig:dPexit}, the exit probability scales as $\theta_{R}^2$ without changing the shape of the distribution, therefore this impacts directly the effective area which follows the same $\theta_{R}^2$ dependency. A comparison with the IceCube effective area is also shown in Fig.~\ref{fig:Aeff} and will be discuss in the following section.

\section{V. Interpretation of the Anomalous Events}

In the previous section we focused our attention on the precise calculation of the ANITA detector effective area, and described how we could obtain a directional probability of detecting an incoming RHN arriving in the direction $(\theta_{\text{DE}},\phi_{\text{RA}})$ and escaping the Earth with an emergence angle $\theta_{\text{em}}$ and an energy ${E}_{\text{exit}}$.

\begin{figure*}
    \centering
    \includegraphics[width=0.495\linewidth]{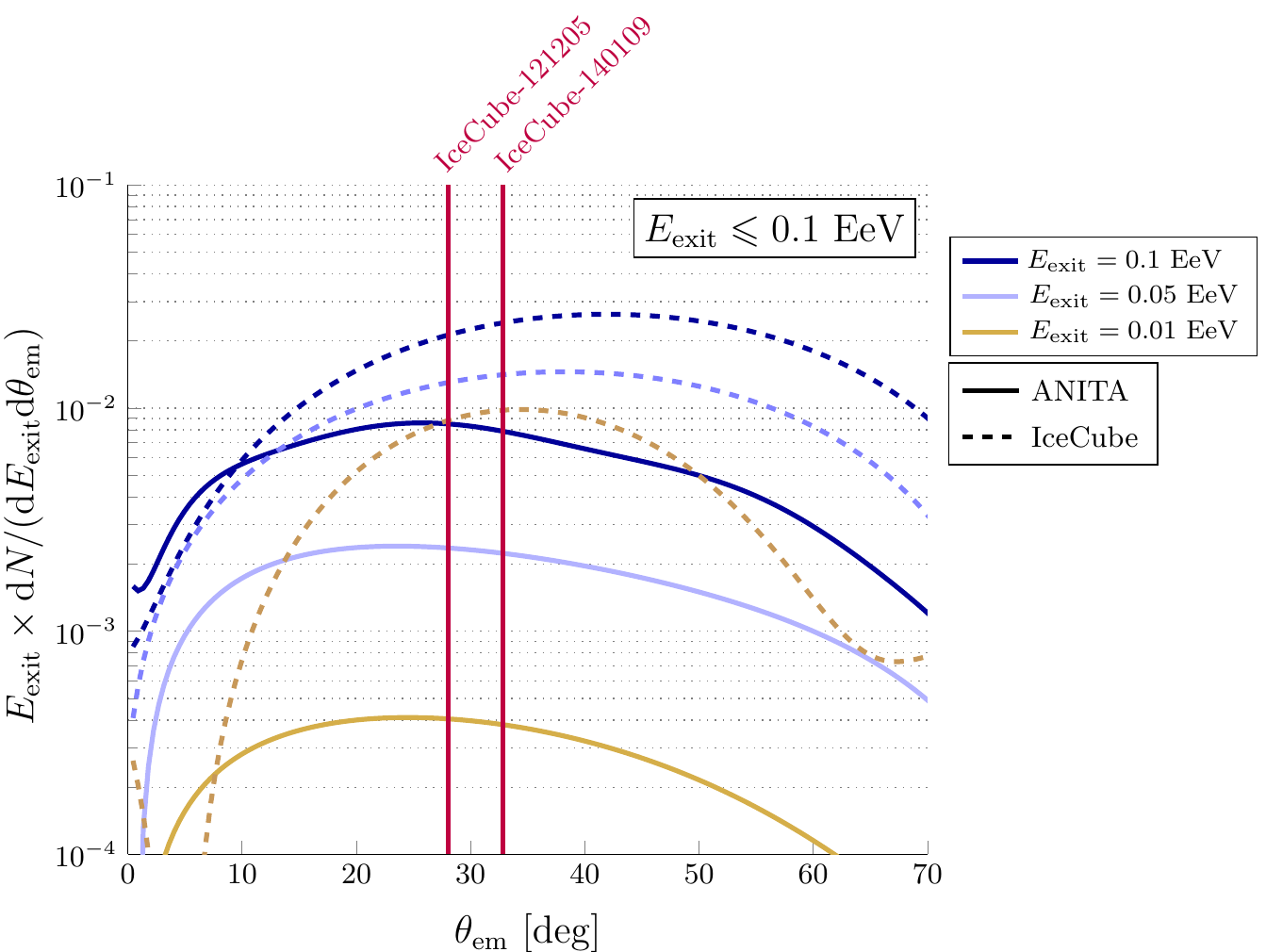}\hspace{0.01\linewidth}\includegraphics[width=0.495\linewidth]{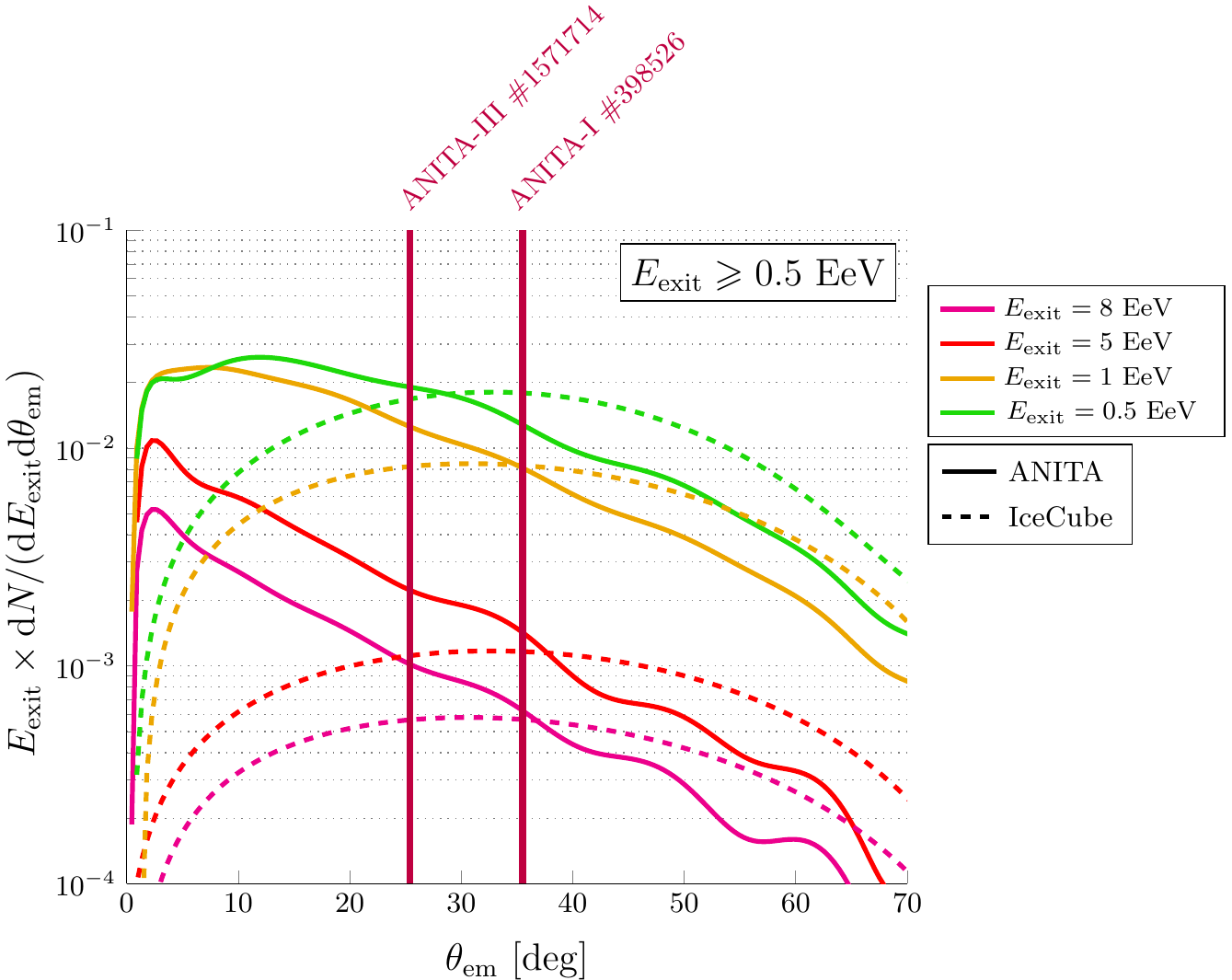}
    \caption{Differential number of events predicted for IceCube and ANITA, per emergence angle $\theta_{\text{em}}$ and exit energy $E_{\text{exit}}$ for an incoming right-handed neutrino energy $E_{N_{R}}=10~\mathrm{EeV}$, a mixing angle $\theta_{R}=0.01$, and a dark-matter lifetime of $10^{23}~\mathrm{s}$.}
    \label{fig:DiffdN_0.01}
\end{figure*}

Given a certain dark-matter mass and a certain dark-matter lifetime, we can use this information to predict the number of events that should have been detected by ANITA for a given energy and emergence angle during its exposure time of $85.5$ days. In the right panel of Fig.~\ref{fig:Aeff} we show the total number of events per emergence angle predicted for ANITA, integrated over all the exit energies for a benchmark point ($m_{\text{DM}}=20$ EeV, $\theta_R$=0.01). Knowing the probability of $\tau$-exit for specific values of the exit energy, we can compute separately the transmission probability for different exit energies and infer the energy range that ANITA is the more sensitive to, given a certain point in the parameter space of our dark-matter model. We present in Fig.~\ref{fig:DiffdN_0.01}
our results for various values of the exit energy $E_{\text{exit}}$ and the same benchmark points than in Fig.~\ref{fig:Aeff} where the number of events predicted for ANITA are depicted by coloured solid lines. Interestingly, and as we already pointed out when discussing the exit probability, one should note that the largest number of events that we predict for ANITA should lie in the energy range $\mathcal{O}(0.5\!-\!1)~\mathrm{EeV}$ and escape the Earth with an emergence angle of order $\mathcal{O}(5^\circ\!-\! 30^\circ)$. Indeed, one can easily see that events of energies $E_\text{exit}\gtrsim 5~\mathrm{EeV}$ and $E_\text{exit}\lesssim 0.1~\mathrm{EeV}$ are less likely to be detected as compared to $\sim$ EeV scale events. This results from the balance of  three different effects: $(i)$ The deep inelastic scattering of neutrinos on nuclei scales with a positive power of the energy. Ultra-high energy neutrinos therefore scatter considerably more than low energy neutrino on Earth producing a larger number of $\tau$-neutrinos deeper under the Earth surface. $(ii)$ The decay length of the taus produced by such scattering scales like the energy due to the boost factor ${E}_{\tau}/m_{\tau}$. Therefore, once produced in the Earth at high energy, such a $\tau$ will propagate and lose energy rather than promptly decay. $(iii)$ Finally the few $\tau$'s that might exit the Earth with a high energy are harder for ANITA to detect since they would not decay within the atmosphere, and would not lead to the production of an EAS.  Such ranges of energies and emergence angles are in relatively good agreement with the features of the two anomalous events observed by ANITA. Furthermore, as far as the total number of events is concerned, for a dark-matter mass of 20~EeV (corresponding to incoming RHN of 10~EeV), a dark-matter lifetime of $\tau_{\text{DM}}=10^{23}\mathrm{s}$ and a mixing angle $\theta_{R}=0.01$, integrating the solid curve in the right panel of Fig.~\ref{fig:Aeff} provides a total number of events of
\begin{equation}
N_\text{tot}\simeq 3.03\text{~events}\,,
\end{equation}
for an exposure of 85.5 days. As a matter of fact, after scanning over the RHN incoming energies $E_{N_{R}}>1~\mathrm{EeV} $, we have derived an interpolation of the expected number of events as function of the parameters of the model in the limit where the mixing angle is relatively small\footnote{Note that for larger values of the mixing angle, the large number of interactions along propagation in the Earth renders the scaling of the number of events with the mixing angle non trivial.}:
\begin{widetext}
\begin{equation}
N_\text{tot}^{\mathrm{ANITA}} \simeq \, 3.03 \, \left(\frac{\theta_R}{0.01} \right)^2
\left( \frac{10^{23}\mathrm{s}}{\tau_\text{DM}} \right)\left( \frac{T_{\mathrm{exp}} }{85.5~\mathrm{days}}\right)
\left( \frac{20 ~\mathrm{EeV}}{m_{\text{DM}}} \right)^{0.67}\qquad \qquad   \left[\ \theta_R\lesssim 0.025\,;~ m_{\text{DM}}>2~\mathrm{EeV}\ \right]\,.  
\label{eq:NtotInterpolation}
\end{equation}
\end{widetext}
Interestingly, asking for a minimum dark-matter lifetime of the order of the age of the Universe ($\tau_{\text{DM}}> 10^{17}$ s), we obtain a minimum value on $\theta_R$ for ANITA to observe $\mathcal{O}(1)$ events : $\theta_R \gtrsim 10^{-5}$. In section III we explored the viable parameter space in the plane $(m_R,\theta_R)$, compatible with the existence of a RHN having the required properties for our setup to work. It is important to stress that regarding the interpretation on the ANITA anomalous events, the RHN mass is not a relevant parameter in our setup, it is expected to be much smaller than its typical kinetic energy $m_\text{DM}/2 \gg m_R$ and behaves as a highly-relativistic particle when it penetrates the Earth.

\subsection{Complementarity with IceCube Detection}

As it was recently pointed out by the ANITA collaboration in the case of the SM neutrino, although IceCube has a significantly smaller effective area compared to ANITA, the fact that it has been running for now almost ten years renders its exposure comparable to ANITA, if not larger for incoming neutrinos of energy $\lesssim 100$~EeV. 
In our case, we saw that neutrinos of energies as large as $\gtrsim 10~\mathrm{EeV}$ could be used to interpret the ANITA anomalous events as being the decay products of a heavy dark-matter particle in the galaxy. It is therefore necessary to understand whether the IceCube collaboration has any reason for not observing events in a similar proportion than ANITA or to exhibit what would be the exact signature of such a decay for IceCube.

Although the IceCube collaboration did not report any event at such high energies, we have seen that $\tau$'s can be produced in the Earth and loose a significant amount of energy while crossing the Earth, which could potentially be detected by IceCube, as long as it would decay within the detector. However, at first approximation, one can notice that a $\sim 20$ PeV $\tau$ has a mean free path of $\sim 1$ kilometer. Above this energy, a $\tau$ behaves (for a $\mathrm{km^3}$ detector like IceCube) as a mere track which is difficult to distinguish from a muon track. In order to estimate the effect of such a decay, we used our very same propagation code to evaluate the amount of events which should have been seen by IceCube over its longest exposure time ($\sim 3142.5~\mathrm{days}$). For this purpose, we performed a Monte-Carlo simulation, demanding that any $\tau$ produced in the Earth decay within a fiducial volume of $\sim 1~\mathrm{km^3}$ located at the South pole and be eventually detected by IceCube. Our results for IceCube are compared to the ANITA effective area shown in Fig.~\ref{fig:Aeff} for an incoming RHN of energy $10~\mathrm{EeV}$ corresponding to a dark-matter particle of mass $m_\text{DM}=20~\mathrm{EeV}$.

As a matter of fact, one can see that IceCube is more sensitive to large emergence angles than ANITA is. Namely, emergence angles larger than $\sim 20^\circ$ are more probable to be detected by IceCube for our choice of benchmark parameters. Such competition was already pointed out in Ref.~\cite{Cherry:2018rxj}. However, we would like to draw the attention of the reader on the fact that such consideration concerns an effective area which is integrated over all the exit-energy range. Since we have seen that the specificity of the $\tau$ trajectory through the Earth can strongly influence its exit energy, it is of interest to compare the probabilities of detection for both ANITA and IceCube exit energy per exit energy. The results of such an analysis is represented in Fig.~\ref{fig:Aeff} (right panel) and Fig.~\ref{fig:DiffdN_0.01} where the differential number of event per emergence angle (integrated over energies in Fig.~\ref{fig:Aeff}) are compared for different values of the exit energy between IceCube and ANITA. Integrating over the emergence angles provides a total number of events predicted for IceCube after $3142.5~\mathrm{days}$ of 3.6 for $\theta_{R}=0.01$ and a dark-matter lifetime $\tau_{\text{DM}}=10^{23}~\mathrm{s}$. Using the same method than for deriving Eq.~(\ref{eq:NtotInterpolation}), we extracted an interpolation of the expected number of events for IceCube:
\begin{widetext}
\begin{equation}
N_\text{tot}^{\mathrm{IceCube}} \simeq \, 3.65 \, \left(\frac{\theta_R}{0.01} \right)^2
\left( \frac{10^{23}\mathrm{s}}{\tau_\text{DM}} \right)\left( \frac{T_{\mathrm{exp}} }{3142.5~\mathrm{days}}\right)
\left( \frac{20 ~\mathrm{EeV}}{m_{\text{DM}}} \right)^{0.70}\qquad \qquad \left[\ \theta_R\lesssim 0.025\,;~ m_{\text{DM}}>2~\mathrm{EeV}\ \right]\,,
\label{eq:NtotInterpolationIceCube}
\end{equation}
\end{widetext}
which is very similar to what ANITA should have seen until now according to Eq.~(\ref{eq:NtotInterpolation}).

Interestingly, it is clear from Fig.~\ref{fig:DiffdN_0.01} that for relatively low energies $\lesssim 0.1~\mathrm{EeV}$, IceCube is more sensitive than ANITA, including at low emergence angles. This is expected since $\tau$'s exiting the Earth with relatively low energy decay close to the surface into an EAS due to their short lifetime. Therefore most of them will decay within the atmosphere and potentially be detected by ANITA. However for such neutrinos to exit the Earth with a low energy, they need to propagate on a sufficiently long distance, so the smaller the emergence angle and the lower the number of events can be detected by ANITA at such energy. Nonetheless, the fiducial volume of IceCube being significantly smaller than the volume of the atmosphere under ANITA, the lower the neutrino energies are, the more the neutrinos can decay inside the detector, and the larger the rate of events is at low energy. This effect of course saturates once $\tau$'s decay with a probability $\sim 1$ inside the detector in the same way this effects saturates for ANITA, but this happens at smaller energies. To put it in a nutshell, EAS produced by $\tau$-leptons exiting the Earth with emergence angles $\lesssim 40^\circ$ are more likely to be detected by ANITA for energies $\gtrsim 0.5~\mathrm{EeV}$, whereas IceCube will become more sensitive than ANITA for EAS of lower energies at similar angles. Given the energy of the anomalous events which have been observed by ANITA --- around $\mathcal{O}(0.5-1)\mathrm{EeV}$ --- it is therefore understandable that such detection has been made by ANITA rather than IceCube.

Interestingly, it was noted in Ref.~\cite{Fox:2018syq} that the anomalous events detected by ANITA could help understanding the IceCube data samples in a better way. According to the authors of Ref.~\cite{Fox:2018syq}, the presence of a sub-EeV neutrino flux could help interpreting the tension between the astrophysical flux required to explain the Northern- versus Southern-hemisphere data samples. In particular they noted that the event IceCube-140611 could possibly be interpreted as coming from an upward-propagating $\tau$, with emergence $\sim 11^\circ$ and energy $\sim 0.07~\mathrm{EeV}$ as was pointed out in Ref.~\cite{Kistler:2016ask}. Moreover the authors of Ref.~\cite{Fox:2018syq} re-analysed the IceCube $\epsilon_\text{proxy} > 200~\mathrm{TeV}$ northern-track sample and could identify two interesting events which could originate from highly-energy up-going neutrinos: IceCube-140109, of energy 0.013~EeV (emergence angle $\sim 32^\circ$) and IceCube-121205 of energy 0.012~EeV (emergence angle $\sim 28^\circ$). If confirmed, such detection from IceCube, in a range of energy $\mathcal{O}(0.01-0.1)~\mathrm{EeV}$ and relatively large emergence angle would match with the complementary prediction that we just pointed out.

Another important feature of our construction is that since the RHN
mixes with the SM active states, for a given dark-matter decay width $\Gamma_\text{DM}$ we expect a subdominant contribution to the dark-matter decay width into SM neutrinos $\Gamma_{\textrm{DM} \rightarrow N_R \nu}= \theta_R^2\Gamma_\textrm{DM}$. 
Therefore we expect a SM neutrino spectral line to reach the surface of the earth with an energy $E_\nu=m_\text{DM}/2$. The IceCube collaboration derived bounds on the DM lifetime based on 
the expected neutrino signal which has not been detected yet~\cite{Aartsen:2018mxl}. As discussed earlier, from Eq.~(\ref{eq:NtotInterpolation}) which is based on our simulation, the total number of events scales
as $N_\text{tot} \propto \tau_\textrm{DM}^{-1} \theta_R^2=\tau_{\textrm{DM} \rightarrow N_R \nu}^{-1}$ 
where $\tau_{\textrm{DM} \rightarrow N_R \nu}^{-1}=\Gamma_{\textrm{DM} \rightarrow N_R \nu}$ is the parameter constrained by the IceCube collaboration. In our case, in order to achieve a number of events of the order of unity $N\sim 1$ this parameter has to be of the order of $\sim 10^{27}$ s which is of the order of the limit derived in~\cite{Aartsen:2018mxl} even though bounds for masses $m_\text{DM}>0.5~\text{EeV}$ are not presented. Therefore, as a complementary feature of our construction, we expected such a neutrino line reaching the Earth to be probed in the following years by the IceCube collaboration.

\section{VI. Extremely-Heavy Dark Matter}

So far we have been focusing on the case of a RHN of incoming energy ${E}_{N_{R}}=10~\mathrm{EeV}$ corresponding to a dark-matter mass $m_\text{DM}=20~\mathrm{EeV}$. In the case of a mixing angle $\theta_{R}=0.01$ we have seen that a dark-matter lifetime $\tau_{\text{DM}}=10^{23}\mathrm{s}$ was appropriate in order to explain why ANITA has seen two anomalous events in 85 days, and possibly why IceCube might have detected a similar number of up-going events at lower energies given its present exposure time. However, the choice of the dark-matter mass (and therefore the RHN incoming energy) was somehow arbitrarily chosen and nothing prevents us, in principle, to consider the case of an even heavier dark-matter particle.

Moreover, we saw earlier that the cosmological constraints coming, including the study of BBN and CMB measurements severely restrict our RHN neutrino to have a mass either smaller then $\lesssim 50~ \mathrm{eV}$ or larger than 50 MeV. In the second case, requiring that our RHN can propagate further than $\sim 50~\mathrm{kpc}$ imposes that our dark-matter candidate has a mass larger than
\begin{equation}
m_\text{DM}\gtrsim 2\times 10^3~\mathrm{EeV}\,.
\end{equation}
We consider here such possibility for completeness and explore to which point it could also provide a possible interpretation to the anomalous events observed by ANITA.

\begin{figure*}
\begin{center}
\includegraphics[width=0.485\linewidth]{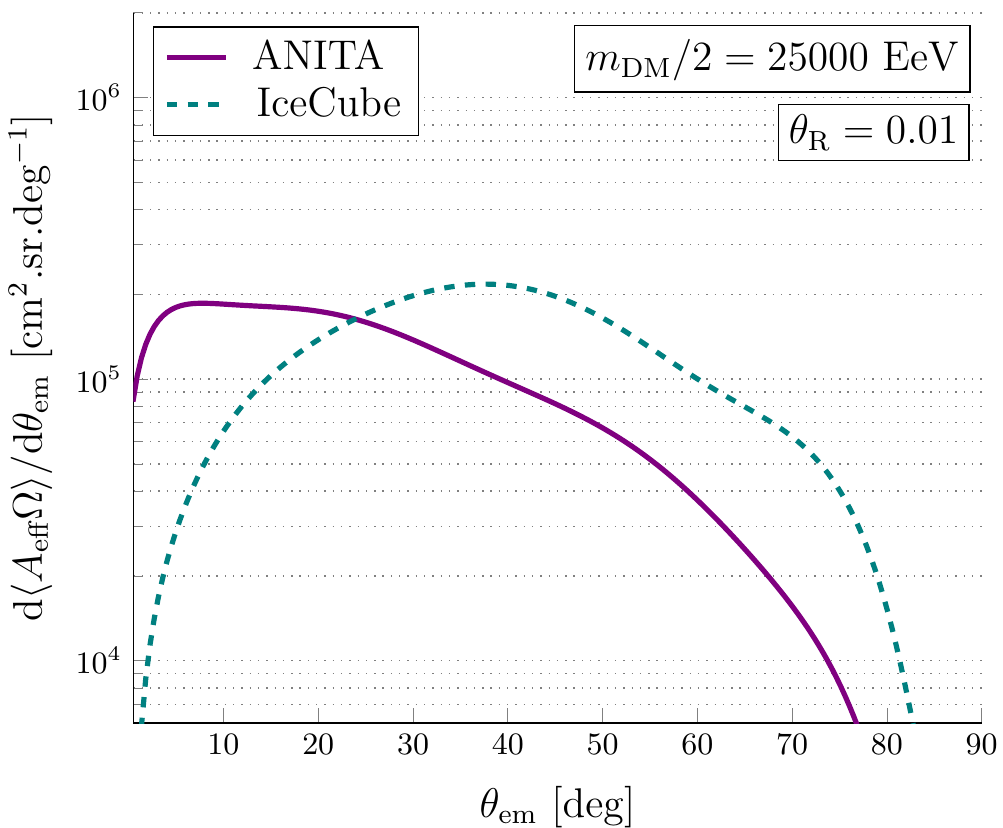}\hspace{0.02\linewidth}\includegraphics[width=0.485\linewidth]{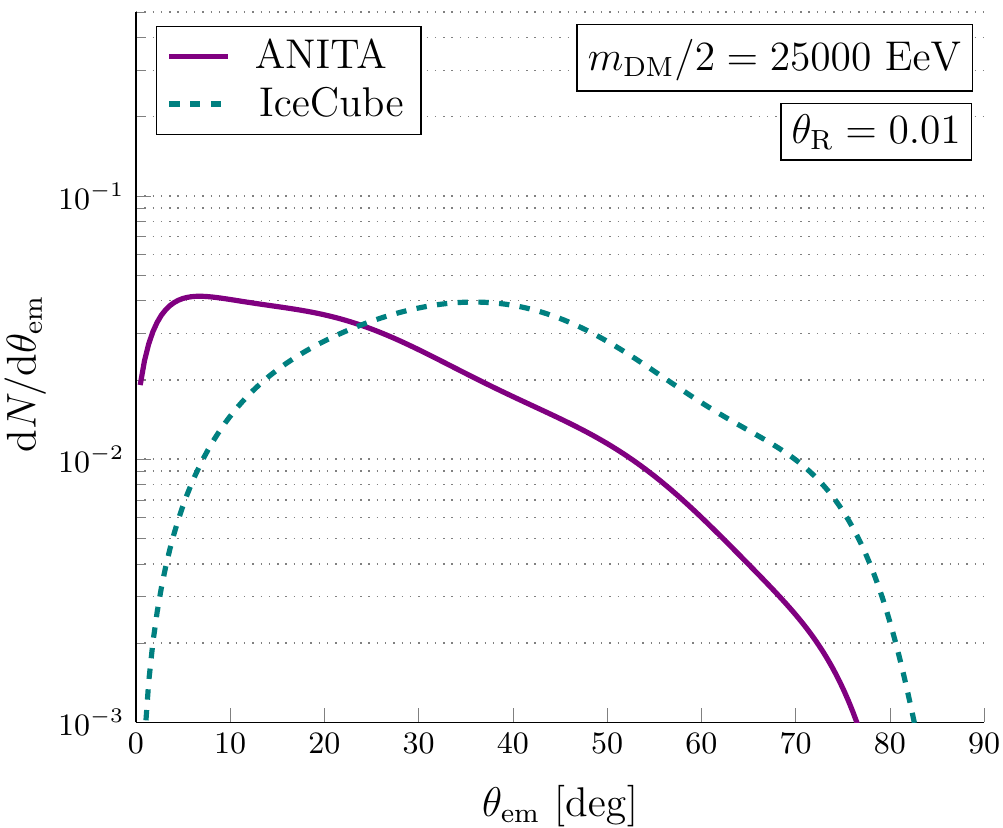} 
\caption{\label{fig:AeffANITAvsIC} Effective area (Left panel) and number of events (Right panel) predicted for IceCube ($T_\text{exp}=3142.5$ days) compared to the one of ANITA ($T_{\text{exp}}=85.5$ days) as a function of the emergence angle, and for an incoming RHN of energy $2.5\times 10^4~\mathrm{EeV}$ and a mixing angle $\theta_{R}=0.01$. In the right panel, the dark-matter lifetime is taken to be $10^{21}~\mathrm{s}$.}
\includegraphics[width=0.49\linewidth]{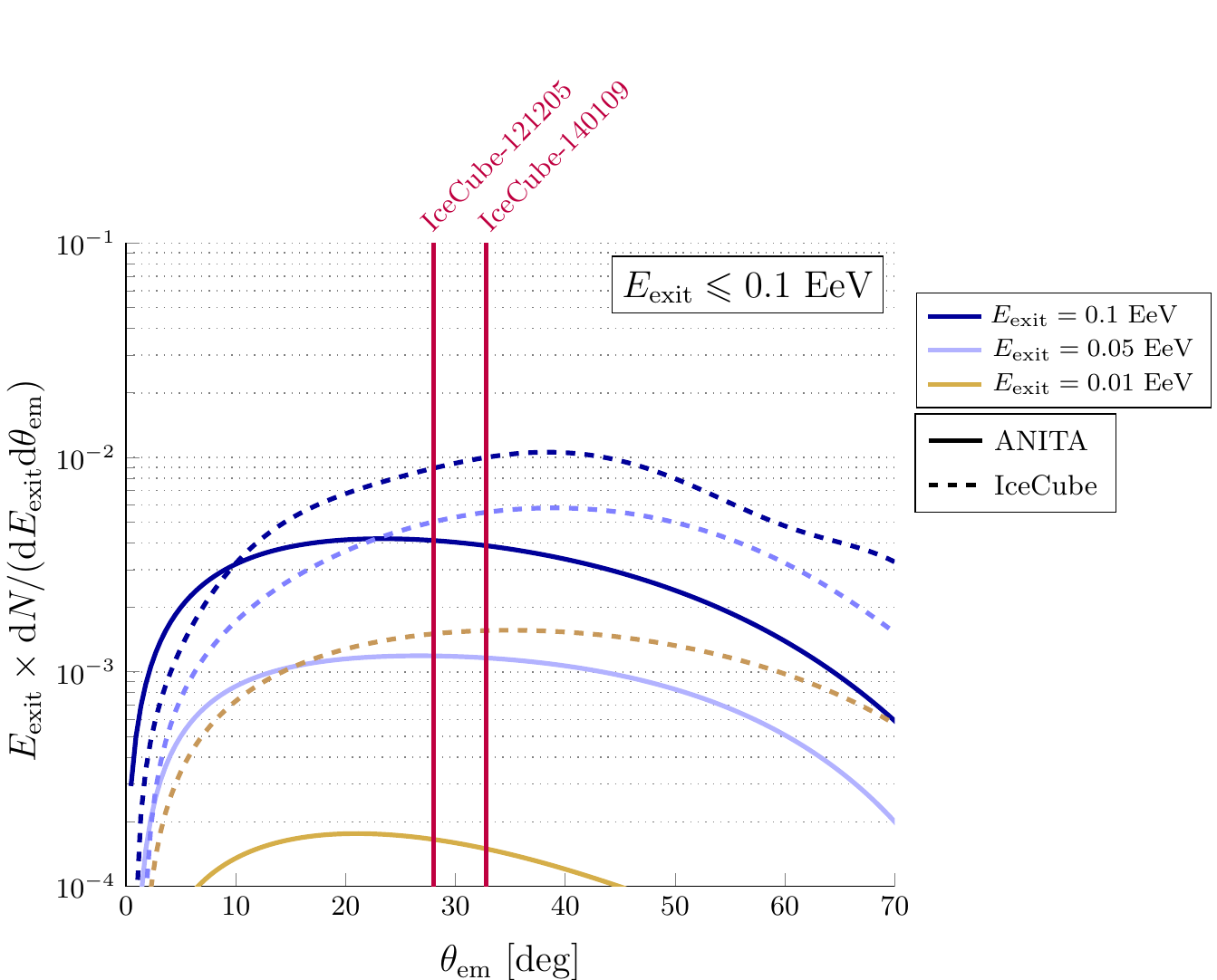}\hspace{0.015\linewidth}\includegraphics[width=0.49\linewidth]{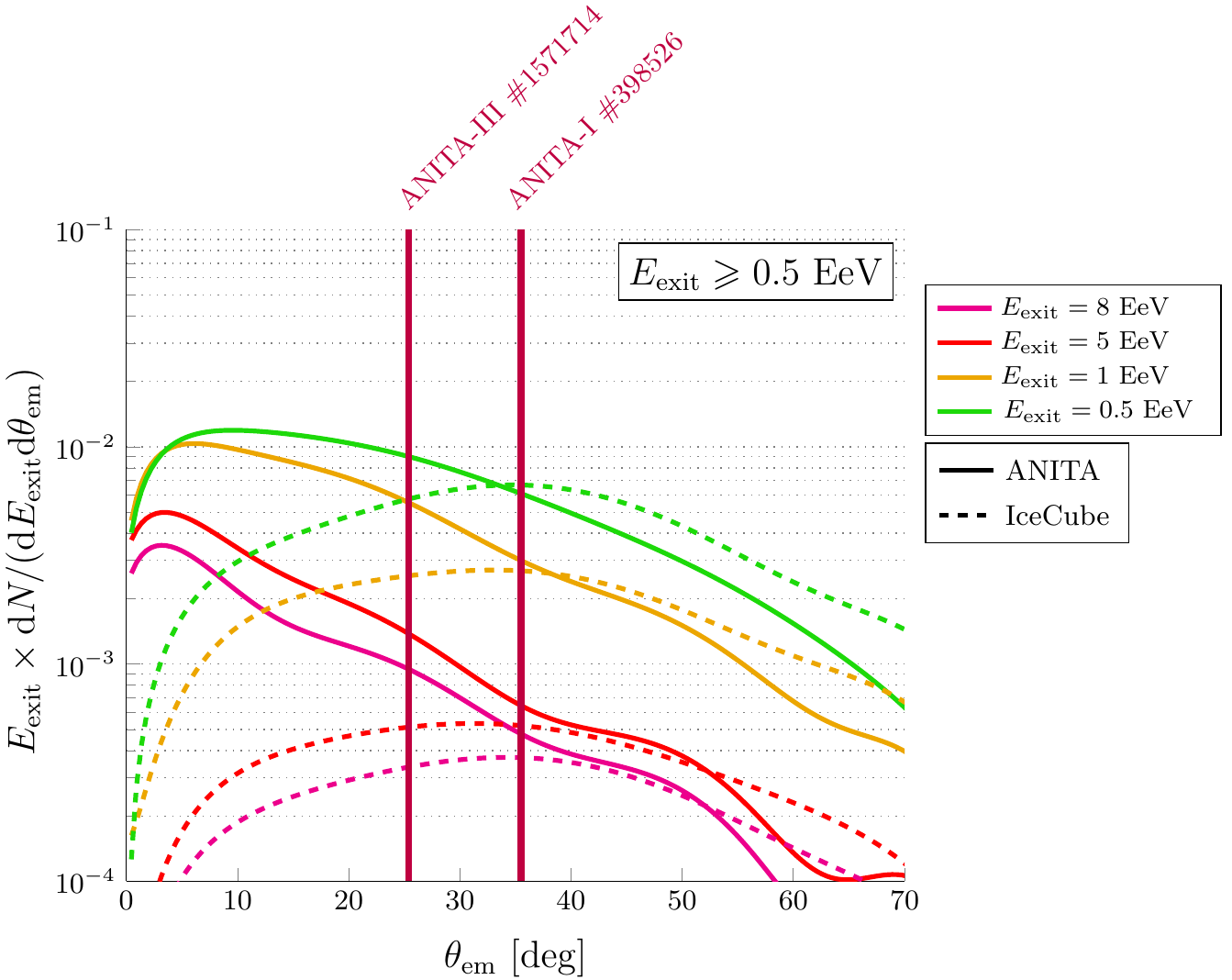} 
\caption{\label{fig:dNdiff_HEAVY} Differential number of events for given exit energies $E_\text{exit}$ predicted for IceCube ($T_{exp}=2431$ days) compared to the one of ANITA ($T_{\text{exp}}=85.5$ days) as a function of the emergence angle, and for an incoming RHN of energy $2.5\times 10^4~\mathrm{EeV}$, a mixing angle $\theta_{R}=0.01$ and a dark-matter lifetime of $10^{21}~\mathrm{s}$.}
\end{center}
\end{figure*}

Using the same propagation code we derived the exit probability per emergence angle and obtain the effective area for both the IceCube and ANITA detectors, as depicted in the left panel of Fig.~\ref{fig:AeffANITAvsIC}. As a matter of fact, the larger the energy of the incoming RHN and the stronger the deep-inelastic scattering interaction is, rendering the Earth less transparent. Therefore we consider the case of a mixing angle $\theta_{R}=0.01$ and note that the distribution shape is relatively unchanged, as compared to the case of a lighter dark-matter particle studied above.

In Fig.~\ref{fig:AeffANITAvsIC} we show the effective area of the detector and the total expected number of events per emergence angle after 85.5 days of operating flight by ANITA in the case of a very heavy dark-matter particle of mass $m_\text{DM}=5\times 10^4~\mathrm{EeV}$ and lifetime $10^{21}~\mathrm{s}$. On this figure one can notice again that the events with a very large energy are less probable as compared to $\sim$ EeV scale events rendering the signature of such a heavy dark-matter particle decay very similar to what we have presented in the previous section. For a lifetime of $\tau_\text{DM}=10^{21}~\mathrm{s}$ and a mixing angle $\theta_{R}=0.01$ we find a total number of events
\begin{equation}
N_\text{tot}\simeq 1.2 \text{~events ~~  (\textbf{ANITA})} 
\end{equation}
for an exposure of 85.5 days. This formula is perfectly in agreement with the expected number for our interpolation Eq.~\eqref{eq:NtotInterpolation}. It is important to notice that from Eq.~\eqref{eq:NtotInterpolation}, for extremely-heavy dark-matter masses, we typically need a smaller dark-matter lifetime to compensate the mass suppression of the generated RHN flux of Eq.~(\ref{eq:flux}) as compared to the $m_\text{DM}$ case.

Similarly to the previous section, a complementary IceCube signal is expected in the extremely-heavy dark matter case. In Fig.~\ref{fig:AeffANITAvsIC} and Fig.~\ref{fig:dNdiff_HEAVY} we show our results for the effective area and expected number of events for the IceCube collaboration after 3142.5  days of data acquisition. The total number of events expected is
\begin{equation}
N_\text{tot}\simeq 1.4 \text{~events ~~  (\textbf{IceCube})}
\end{equation}
therefore predicting the same agreement between the number of events observed by ANITA and IceCube than in the case of a $\mathcal \sim 10~\mathrm{EeV}$ dark-matter particle.

Moreover, it was shown in~\cite{Berezinsky:2002hq} that SM neutrino production from DM decay associated to on-shell production of electroweak gauge bosons is comparable to the production of SM neutrinos only case\footnote{The main argument is that a strong enhancement from large logarithms would compensate the smallness of the extra gauge coupling involved in the diagram with an extra gauge boson in the final state.}. Based on this argument, the dark-matter decay is expected to be associated with the production of gauge bosons therefore inducing so-called Electro-Weak Lepton-Boson (EWLB) cascades at lower energies. Such cascades would induce a continuous spectrum of SM charged particles and high energy neutrinos that could be observable by IceCube. Such an idea was used by the authors of~\cite{Kachelriess:2018rty} to derive bounds on  super-heavy dark-matter partial decay width to neutrinos which excludes values of $\tau_{\textrm{DM} \rightarrow N_R \nu}$ higher than $\sim 10^{27}$ s for $m_\text{DM}\sim10^4~\text{EeV}$. As stressed out before, in this case, we expect the DM decay lifetime to be typically smaller than in the $m_\text{DM}\sim1~\text{EeV}$ case in order to account for an ANITA expected number of events $N \sim 1$. Therefore, even though physics at the $10^4~\text{EeV}$ scale still remains quite speculative up to this day, we expect the extremely-heavy dark matter ANITA interpretation to be in conflict with the IceCube bound from~\cite{Kachelriess:2018rty} by 1-2 orders of magnitude. Note however that the presence of new physics, such as the mixing with additional non-SM states which are sufficiently long-lived, might alter significantly the discussion presented in~\cite{Kachelriess:2018rty}. Such study is out of the scope of this paper, therefore we leave such investigations for future work.

\section{Conclusion}

In this paper we have studied the possibility that the two anomalous events observed by ANITA can be interpreted as an indirect signal of the presence of dark-matter particles in the galaxy. We have considered the simple case of a heavy dark-matter candidate decaying in the Milky-Way halo into a pair of right-handed neutrinos, mixed with the Standard Model active neutrinos. 
We have shown that for such a right-handed neutrino to propagate on sufficiently large distances through the galaxy while satisfying constraints from BBN, and mixing with neutrinos with a mixing angle smaller than 1, the right-handed neutrino could in principle lie in two different mass ranges, namely of $\mathcal{O}(1-10)~\mathrm{eV}$ ($\mathcal{O}(0.01-1)~\mathrm{GeV}$) and boosted with energies as large as $10~\mathrm{EeV}$ ($10^{4}~\mathrm{EeV}$). {Interestingly, the viable right-handed-neutrino mass range points towards regions where such a state might be used for successful low-scale leptogenesis, to alleviate tensions between cosmological observations or to address discrepancies in neutrino short-baseline experiments.}

We have adapted the Monte-Carlo simulation code provided with Ref.~\cite{Alvarez-Muniz:2017mpk} in order to simulate accurately the neutrino production, propagation and conversion into $\tau$-leptons through the Earth. We could analyze in details the probability for a $\tau$-lepton of arbitrary energy ${E}_\text{exit}$ to exit the Earth, after being
generated by a incoming right-handed neutrino of energy ${E}_N=m_\text{DM}/2$ with a specific direction crossing the Earth and converting into Standard Model states.

We showed that for a dark-matter particle of mass $20~\mathrm{EeV}$ and $5\!\times\! 10^{4}~\mathrm{EeV}$, and of respective lifetimes $\theta_R^2\!\times\!10^{27}~\mathrm{s}$ and $ \theta_R^2\!\times\!10^{25}~\mathrm{s}$, decaying into a pair of RHN mixing with SM neutrinos with mixing angle $\theta_{R}$, the ANITA collaboration should have reported $\sim \mathcal{O}(1)$ events after 85 days of operating flight in the energy range $\mathcal{O}(0.5-1)~\mathrm{EeV}$.

We have also studied the possibility of a detection by the IceCube experiment, which, due to its very large exposure time, is competitive with ANITA to detect such ultra-high energy neutrino cosmic rays, as was already noticed in Ref.~\cite{Cherry:2018rxj,Huang:2018als}. We used the very same propagation code in order to simulate such detection {and we derived the following interpolation formulas for the expected number of events based on our simulation:
\begin{widetext}
\begin{align*}
N_\text{tot}^{\mathrm{ANITA}} \simeq& \, 3.03 \, \left(\frac{\theta_R}{0.01} \right)^2
\left( \frac{10^{23}\mathrm{s}}{\tau_\text{DM}} \right)\left( \frac{T_{\mathrm{exp}} }{85.5~\mathrm{days}}\right)
\left( \frac{20 ~\mathrm{EeV}}{m_{\text{DM}}} \right)^{0.67}\qquad \qquad  &\left[\ \theta_R\lesssim 0.025\,;~ m_{\text{DM}}>2~\mathrm{EeV}\ \right]\,,
\\
N_\text{tot}^{\mathrm{IceCube}} \simeq& \, 3.65 \, \left(\frac{\theta_R}{0.01} \right)^2
\left( \frac{10^{23}\mathrm{s}}{\tau_\text{DM}} \right)\left( \frac{T_{\mathrm{exp}} }{3142.5~\mathrm{days}}\right)
\left( \frac{20 ~\mathrm{EeV}}{m_{\text{DM}}} \right)^{0.70}\qquad \qquad &\left[\ \theta_R\lesssim 0.025\,;~ m_{\text{DM}}>2~\mathrm{EeV}\ \right]\,.
\label{eq:NtotInterpolation}
\end{align*}
\end{widetext}
}
We confirmed that both experiments should expect detecting a very similar number of events given their present exposure. Moreover we analysed the spectrum of events which should have been detected by IceCube and we found out that events of energies around $\mathcal{O}(0.5-1)~\mathrm{EeV}$ are more likely to be seen by ANITA at emergence angles $\mathcal{O}(0^\circ-35^\circ)$ while IceCube would start being competitive with ANITA below $30^\circ$ only for event energies as small as $\lesssim 0.1~\mathrm{EeV}$. Interestingly such conclusion is in very good agreement with the proposal of Ref.~\cite{Kistler:2016ask} which showed that the IceCube collaboration might have already detected three events of this kind in their data samples without reporting it.

As a refreshing feature of the case of extremely heavy dark-matter, we would like to emphasize that the region of the parameter space where our RHN neutrino is as heavy as $0.1~\mathrm{GeV}$ favours models with a dark-matter mass around $\gtrsim 5\times 10^{13}~\mathrm{GeV}$ in order to $(i)$ escape the cosmological bounds on RHN, $(ii)$ be able to interpret the ANITA events and $(iii)$ ensure that our dark-matter candidate is stable on cosmological time scales. One should note that such a large mass value is surprisingly matching with the mass scale required for the inflaton particle in most of the large-field scenarios of cosmic inflation. The possibility that the inflaton particle itself constitutes the main matter component of our universe is something which gained interest in the last decade \cite{Hooper:2018buz,Almeida:2018oid,Rosa:2018iff,Lidsey:2001nj,Barshay:1998ys,Barshay:1997ni, Borah:2018rca} and which could potentially arise if, for some reason, the inflaton is not able to decay anymore after it reheats significantly the universe. If a dark-matter particle is proven in the next years to be detected through its decay products by ANITA and IceCube at such high energies, the connection between inflationary physics and dark-matter phenomenology would therefore turn out to be among the most challenging topics of phenomenological studies in the future.\\

Finally, more data are to be accumulated by ANITA which already realized its fourth flight and should reveal their most recent results in a few months, which might bring more insights about the morphology of the signal and in particular the statistically favoured emergence angle necessary to interpret accurately the signal. \\

\noindent {\bf Acknowledgements. } 
The authors would like to thank Peter Gorham and Andrew Romero-Wolf for their useful comments about the ANITA experiment and its detection efficiency as well as K.A. Olive, Fei Huang and Doojin Kim for fruitful discussions. MP would like to thank Enrique Fernández-Martinez, Jacobo López-Pavón, Olga Mena, Stefano Gariazzo and Vivian Poulin for fruitful conversations.
The research activities of LH are supported  by the Department of Energy under Grant DE-FG02-13ER41976/DE-SC0009913 and by the CNRS. 
YM is supported by the France-US PICS MicroDark and acknowledges
partial support from the European Union Horizon 2020 research and
innovation programme under the Marie Sklodowska-Curie: RISE
InvisiblesPlus (grant agreement No 690575) and the ITN Elusives (grant
agreement No 674896).
The work of MP was supported by the Spanish Agencia Estatal de Investigaci\'{o}n through the grants FPA2015-65929-P (MINECO/FEDER, UE), IFT Centro de Excelencia Severo Ochoa SEV-2016-0597, and Red Consolider MultiDark FPA2017-90566-REDC. LH would like to thank the CPHT of \'Ecole Polytechnique for its hospitality during part of the realization of this work.

\end{document}